\documentclass{intech} 

\booktitle{Will-be-set-by-IN-TECH}%

\chaptertitle{A Microscopic Equation of State for Neutron-Rich Matter and its Effect on   
Neutron Star Properties} 

\authors{Francesca Sammarruca} 
\affiliation{University of Idaho}
\country{USA}

\begin{document}

\maketitle

\section{Introduction }
In this chapter, we will be concerned with 
hadronic interactions
in the nuclear medium, particularly under conditions of extreme densities such as those 
encountered in some compact astrophysical objects. 
This issue  goes to the very core 
of nuclear physics. In fact, our present knowledge of the nuclear force in free space is, in itself, 
the result of decades of struggle \citep{Mac89} which will not be reviewed in this chapter.  
The nature of the nuclear force in the medium is of course an even more complex problem, 
as it involves aspects of the force that cannot be constrained
through free-space nucleon-nucleon (NN) scattering. Predictions of properties of nuclei are 
the ultimate test for many-body theories. 

Nuclear matter is a convenient theoretical laboratory for many-body theories. By "nuclear matter" we mean an infinite system 
of nucleons acted on by their mutual strong forces and no electromagnetic interactions. Nuclear matter 
is characterized by its energy per particle as a function of density and other thermodynamic quantities, as  
appropriate (e.g.~temperature). Such relation is known as the nuclear matter equation of state (EoS). 
The translational invariance of the system facilitates theoretical calculations. At the same time, adopting     
what is known as 
the "local density approximation", one can use the EoS to obtain information on finite systems. This procedure 
is applied,   
for instance, in Thomas-Fermi calculations within the liquid drop model, where an appropriate energy functional is 
written in terms of the EoS \citep{Oya98,Furn,SL09}.

Isospin-asymmetric nuclear matter (IANM) simulates the interior of a nucleus with unequal densities of protons and neutrons.
The equation of state of (cold) IANM is then a function of density as well as the relative concentrations 
of protons and neutrons. 

The recent and fast-growing interest in IANM stems from its close connection to the physics of neutron-rich nuclei, or,
more generally, isospin-asymmetric nuclei, including the very "exotic" ones known as "halo" nuclei. 
At this time, the boundaries of the nuclear chart are uncertain, with several                          
thousands nuclei believed to exist but not yet observed in terrestrial laboratories. 
The Facility for Rare Isotope Beams (FRIB) has recently been approved for design and construction at 
Michigan State University.             
The facility will deliver intense beams of rare isotopes, the study of which can provide crucial 
information on short-lived elements normally not found on earth.                  
Thus, this new experimental program will have widespread impact, ranging from the origin of elements to the              
evolution of the cosmos. 
It is estimated that 
the design and construction of FRIB will take ten years.                 
In the meantime, systematic investigations to determine
the properties of asymmetric nuclear matter are proliferating at existing facilities. 

The equation of state of IANM is also the crucial input for the structure equations of
compact stars, and thus establishes the connection between nuclear physics and compact astrophysical systems. 
It is the focal point of this chapter to present and discuss our approach to the 
devolopment of the EoS of nuclear and neutron-rich matter, with particular 
emphasis on the latter and its relation to the properties of neutron stars. 

The chapter will articulate through the following sections: In Section {\bf 2}, we present a 
brief review of facts and phenomenology about IANM. We then proceed to describe our
microscopic approach to calculate the energy per particle in IANM (Section {\bf 3}) and show 
the corresponding predictions. Section {\bf 4} will be dedicated to a review of neutron
star structure and available empirical constraints. Microscopic predictions of neutron star 
properties will be presented and discussed there. Section {\bf 5} contains a brief discussion 
on the topic of polarized IANM. 
 The chapter will end with our 
conclusions and an outlook into the future (Section {\bf 6}).

\section{Facts about isospin-asymmetric nuclear matter} 

Asymmetric nuclear matter can be characterized by the neutron density, 
$\rho_n$, and the proton density, $\rho_p$, defined as the number of neutrons or protons per unit of volume. 
In infinite matter, they are obtained by summing the neutron or proton states per volume (up to their respective 
Fermi momenta, $k^{n}_{F}$ or $k^{p}_{F}$) and applying the appropriate degeneracy factor. The result is 
\begin{equation}
  \rho_i =\frac{ (k^{i}_{F})^3}{3 \pi ^2} ,   \label{rhonp}   
\end{equation}
with $i=n$ or $p$. 

It may be more convenient to refer to the total density
$\rho = \rho_n + \rho_p$ and the asymmetry (or neutron excess) parameter
$\alpha = \frac{ \rho_n - \rho_p}{\rho}$. 
Clearly, $\alpha$=0 corresponds to symmetric matter and 
$\alpha$=1 to neutron matter.                       
In terms of $\alpha$ and the average Fermi momentum, $k_F$, related to the total density in the usual way, 
\begin{equation}
  \rho =\frac{2 k_F^3}{3 \pi ^2} ,   \label{rho}   
\end{equation}
the neutron and proton Fermi momenta can be expressed as 
\begin{equation}
 k^{n}_{F} = k_F{(1 + \alpha)}^{1/3}            \label{kfn}
\end{equation}
and 
\begin{equation}
 k^{p}_{F} = k_F{(1 - \alpha)}^{1/3} ,            \label{kfp} 
\end{equation}
 respectively.

Expanding 
the energy per particle in IANM  with respect to the asymmetry parameter yields
\begin{equation}
e(\rho, \alpha) = e_0({\rho}) + \frac{1}{2} \Big (\frac{\partial ^2 e(\rho,\alpha)}{\partial \alpha ^2}\Big )_{\alpha=0}\alpha ^2 +{\cal O}(\alpha ^4) \; , \label{exp}  
\end{equation}
where the first term is the energy per particle in symmetric matter and 
the coefficient of the quadratic term is identified with the symmetry energy,
$e_{sym}$. In the Bethe-Weizs{\" a}cker formula for the nuclear binding energy, it represents the amount of binding a nucleus has 
to lose when the numbers of protons and neutrons are unequal.                                             
A typical value for $e_{sym}$               
at nuclear matter density ($\rho_0$) is 30 MeV, 
with theoretical predictions spreading approximately between 26 and 35 MeV.

To a very good degree of approximation, 
the energy per particle in IANM can be written as 
\begin{equation}
e(\rho, \alpha) \approx e_0({\rho}) + e_{sym}(\rho)\alpha ^2.   \label{e}                    
\end{equation} 
The effect of a term of fourth order in the asymmetry parameter (${\cal O}(\alpha ^4)$) on the bulk properties of neutron stars 
is very small, although it may impact the proton fraction at high density. 

Equation~(\ref{e}) displays a convenient separation between the symmetric and aymmetric parts of the EoS, 
which facilitates the identification of observables that may be sensitive, for instance, mainly to the 
symmetry energy. At this time, groups from GSI \citep{GSI1,GSI2}, MSU \citep{Tsang}, Italy \citep{Greco}, France, \citep{IPN},          
China \citep{China1,China2}, and Japan \citep{RIKEN} 
are investigating the density dependence of the symmetry energy through heavy-ion collisions. 
Typically, constraints are extracted from heavy-ion collision simulations 
based on transport models. 
Isospin diffusion and the ratio of neutron and proton spectra are among the 
observables used in these analyses. 

These investigations appear to agree reasonably well on the following parametrization 
of the symmetry energy: 
\begin{equation}
e_{sym}(\rho) = 12.5 \, MeV \Big (\frac{\rho}{\rho_0}\Big )^{2/3} +                       
17.5 \, MeV \Big (\frac{\rho}{\rho_0}\Big )^{\gamma_i},                   \label{es} 
\end{equation} 
where $\rho_0$ is the saturation density. 
The first term is the kinetic contribution and 
 $\gamma_i$ (the exponent appearing in the potential energy part) is found to be between 0.4 and 1.0. 
Recent measurements of elliptic flows in $^{197}$Au + $^{197}$Au reactions at GSI  at  
400-800 MeV per nucleon 
favor a potential energy term with $\gamma_i$ equal to 0.9 $\pm$ 0.4.           
Giant dipole resonance excitation in fusion reactions \citep{GDR} is also sensitive to
the symmetry energy, since the latter is responsible for isospin equilibration 
in isospin-asymmetric collisions.

Isospin-sensitive observables can also be identified among the properties of normal nuclei. 
The neutron skin of neutron-rich nuclei is a powerful isovector observable, being sensitive to the   
slope of the symmetry energy, which determines to which extent neutrons will tend to spread outwards 
 to form the skin.

Parity-violating electron scattering experiments are now a realistic option        
to determine neutron distributions with unprecedented accuracy. The neutron radius of 
$^{208}$Pb is expected to be measured with a precision of 3\% thanks to the electroweak program
at the Jefferson Laboratory, the PREX experiment in particular, just recently completed at Jefferson Lab. This level of accuracy could not be achieved with hadronic scattering. 
Parity-violating electron scattering at low momentum transfer is especially suitable to probe neutron densities, as the     
 $Z^0$ boson couples primarily to neutrons. 
With the success of this program, 
 reliable empirical information on neutron skins will be able to provide, in turn, much needed {\it independent} constraint on the 
density dependence of the symmetry energy.

A measure of the density dependence of the symmetry energy is 
 the symmetry pressure, defined as 
\begin{equation}
L = 3 \rho_0 \Big (\frac{\partial e_{sym}(\rho)}{\partial \rho}\Big )_{\rho_0} \approx 
 3 \rho_0 \Big (\frac{\partial e_{n.m.}(\rho)}{\partial \rho}\Big )_{\rho_0} \, , 
\label{L} 
\end{equation} 
where we have used Eq.~(\ref{e}) with $\alpha$=1. 
Thus, $L$ is sensitive to the gradient of the energy per particle in neutron matter ($e_{n.m.}$). 
As to be expected on physical grounds, the neutron skin, given by                                  
\begin{equation}
S = \sqrt{<r_n^2>} - \sqrt{<r_p^2>} \, \, , 
\label{S} 
\end{equation} 
is highly sensitive to the same pressure gradient.

Values of $L$ are reported to range 
from -50 to 100 MeV as seen, for instance, through the numerous
parametrizations of Skyrme interactions,           
 all chosen to fit the binding energies and the 
charge radii of a large number of nuclei   
(\citep{BA05} and references therein).              
Heavy-ion data impose boundaries for $L$ at $85 \pm 25$ MeV, with more              
stringent constraints being presently extracted.                    
At this time constraints appear to favor lower values of the symmetry pressure.       
In fact, a range of $L$ values given by  
$52.7 \pm 22.5$ MeV has emerged from recent analyses of global optical potentials        
\citep{BA10}. 

Typically, parametrizations like the one given in Eq.~(\ref{es}) are valid 
at or below the saturation density, $\rho_0$. Efforts to constrain the behavior of the symmetry energy
at higher densities 
are presently being pursued through observables such as $\pi ^-/\pi^+$ ratio, 
$K ^+/K^0$ ratio, neutron/proton differential transverse flow, or nucleon elliptic flow \citep{Ko09}. 

Another important quantity which emerges from studies of IANM is the symmetry potential.
Its definition stems from the observation that the single-particle potentials 
experienced by the proton and the neutron in IANM, $U_{n/p}$, are different from each other and satisfy      
the approximate relation 
\begin{equation}
U_{n/p}(k,\rho,\alpha) \approx U_{n/p}(k,\rho,\alpha=0) \pm U_{sym}(k,\rho)\;\alpha \; , 
\label{Unp}
\end{equation}
where the +(-) sign refers to neutrons (protons), and               
\begin{equation}
U_{sym}=\frac{U_{n} - U_p}{2\alpha} \; .                  
\label{Usym} 
\end{equation}
(Later in the chapter we will verify the approximate linear behavior with 
respect to $\alpha$ displayed in Eq.~(\ref{Unp}).)
Thus, 
one can expect isospin splitting of the single-particle potential to be effective in separating           
the collision dynamics of neutrons and protons. 
Furthermore, $U_{sym}$, being proportional to the gradient between the single-neutron and
the single-proton potentials, 
should be comparable with the Lane potential \citep{Lane}, namely the isovector 
part of the nuclear optical potential. Optical potential analyses can then 
help constrain this quantity and, in turn, the symmetry energy.           

Because of the fundamental importance of the symmetry energy in many systems and phenomena,
it is of interest to identify 
 the main contributions to its density dependence.                                
In a recent work \citep{FS11} we discussed the 
contribution of the isovector mesons ($\pi$, $\rho$, and $\delta$) to the 
symmetry energy and demonstrated the chief role of the pion. Note that the 
isovector mesons carry the isospin dependence by contributing differently in different
partial waves, and that isospin dependence controls the physics
of IANM. 
Hence, we stress the relevance of a microscopic model that contains all
important couplings of mesons with nucleons. 

\section{Our microscopic approach to isospin-asymmetric nuclear matter}
\subsection{The two-body potential} 

Our approach is
{\it ab initio} in that the starting point of the many-body calculation is a realistic NN interaction which is then applied in the 
nuclear medium without any additional free parameters. 
Thus the first question to be confronted concerns the choice of the "best" NN interaction. 
After the development of Quantum Chromodynamics (QCD) and the understanding of its symmetries,  
chiral effective theories \citep{chi} were developed as a way to respect the 
symmetries of QCD while keeping the degrees of freedom (nucleons and pions) typical of low-energy nuclear physics. However, 
chiral perturbation theory (ChPT)
has definite limitations as far as the range of allowed momenta is concerned. 
For the purpose of applications in dense matter, where higher and higher momenta become involved     
with increasing Fermi momentum, NN potentials based on ChPT are unsuitable.       

Relativistic meson theory is an appropriate framework to deal with the high momenta encountered in dense
matter. In particular, 
the one-boson-exchange (OBE) model has proven very successful in describing NN data in free space 
and has a good theoretical foundation. 
Among the many available OBE potentials, some being part of the "high-precision generation" \citep{pot1,N93,V18}, 
we seek a momentum-space potential developed within a relativistic scattering equation, such as the 
one obtained through the Thompson \citep{Thom} three-dimensional reduction of the Bethe-Salpeter equation \citep{BS}. 
Furthermore, we require a potential that uses 
the pseudovector coupling for the interaction of nucleons with pseudoscalar mesons. 
With these constraints in mind, 
as well as the requirement of a good description of the NN data, 
Bonn B \citep{Mac89} is a reasonable choice. As is well known, the NN potential model dependence
of nuclear matter predictions is not negligible. The saturation points obtained with different NN potentials
move along the famous "Coester band" depending on the strength of the tensor force, with the weakest tensor
force yielding the largest attraction. This can be understood in terms of medium effects (particularly 
Pauli blocking) reducing the (attractive) second-order term in the expansion of the reaction matrix. 
A large second-order term will undergo a large reduction in the medium. Therefore, noticing that the second-order term
is dominated by the tensor component of  the force, nuclear potentials with a strong tensor component will
yield less attraction in the medium. 
For the same reason (that is, the role of the tensor force in  
nuclear matter), 
the potential model dependence is strongly reduced in pure (or nearly pure) neutron matter, due to the  
absence of isospin-zero partial waves. 

\begin{figure}
\centering            
\vspace*{-3.2cm}
\hspace*{-1.5cm}
\scalebox{1.0}{\includegraphics{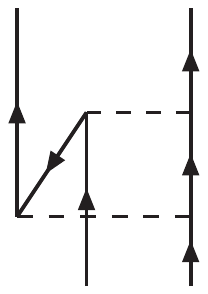}}
\vspace*{-21.0cm}
\caption{Contribution to the NN interaction from virtual pair excitation.                   
Upward- and downward-pointing arrows represent nucleons and antinucleons, respectively.
Dashed lines denote mesons.                            
} 
\label{2b}
\end{figure}

Already when QCD (and its symmetries) were unknown, it was observed that the contribution from the
nucleon-antinucleon pair diagram, Fig.~\ref{2b}, becomes unreasonably large if the pseudoscalar (ps) coupling is used, 
leading to very large pion-nucleon scattering lengths \citep{GB79}.                                            
We recall that the Lagrangian density for pseudoscalar coupling of the nucleon field ($\psi$) with the  pseudoscalar meson
field ($\phi$) is 
\begin{equation}
{\cal L}_{ps} = -ig_{ps}\bar {\psi} \gamma _5 \psi \phi.     \label{ps} 
\end{equation} 
On the other hand, the same contribution, shown in Fig.~\ref{2b}, 
is heavily reduced by the pseudovector (pv) coupling (a mechanism which
became known as "pair suppression"). The reason for the suppression is the presence of the 
covariant derivative                                                                                     
at the pseudovector vertex,                                                  
\begin{equation}
{\cal L}_{pv} = \frac{f_{ps}}{m_{ps}}{\bar \psi}  \gamma _5 \gamma^{\mu}\psi \partial_{\mu} \phi,              
\label{pv} 
\end{equation} 
which reduces the contribution of the vertex for low momenta and, thus, 
 explains the small value of the pion-nucleon
scattering length at threshold \citep{GB79}. 
Considerations based on chiral symmetry \citep{GB79} can further motivate 
the choice of the pseudovector coupling.                          

\begin{figure}[!t] 
\centering         
\vspace*{-3.2cm}
\hspace*{-1.0cm}
\scalebox{0.9}{\includegraphics{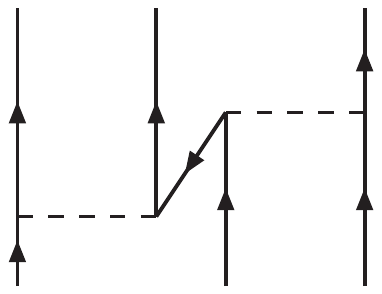}}
\vspace*{-19.0cm}
\caption{Three-body force due to virtual pair excitation. Conventions as in the previous figure.
} 
\label{3b}
\end{figure}

In closing this section, we wish to highlight          
the most important aspect of the {\it ab initio} approach: Namely, the only free parameters of the
model (the parameters of the NN potential)                                               
are determined by fitting the free-space NN data and never readjusted in the medium. In other
words, the model parameters are tightly constrained and the calculation in the medium is 
parameter free. 
The presence of free parameters in the medium would generate effects and sensitivities which are hard to
control and interfere with the predictive power of the theory. 

\subsection{The Dirac-Brueckner-Hartree-Fock approach to symmetric and  asymmetric nuclear matter}

\subsubsection{Formalism} 

The main strength of the DBHF approach is its inherent ability to account for important three-body forces   
through its density dependence. 
In Fig.~\ref{3b} we show a three-body force (TBF) originating from virtual excitation of a nucleon-antinucleon pair, 
known as "Z-diagram". Notice that the observations from the previous section ensure that the corresponding diagram
at the two-body level, Fig.~\ref{2b}, is moderate in size when the pv coupling,           
Eq.~\ref{pv}, is used. 
The main feature of                                     
the DBHF method turns out to be closely related to 
the TBF depicted in Fig.~\ref{3b}, as we will argue next. In the DBHF approach, one describes the positive energy solutions
of the Dirac equation in the medium as 
\begin{equation}
u^*(p,\lambda) = \left (\frac{E^*_p+m^*}{2m^*}\right )^{1/2}
\left( \begin{array}{c}
 {\bf 1} \\
\frac{\sigma \cdot \vec {p}}{E^*_p+m^*} 
\end{array} 
\right) \;
\chi_{\lambda},
\label{ustar}
\end{equation}
where the nucleon effective mass, $m^*$, is defined as $m^* = m+U_S$, with $U_S$ an attractive scalar potential.
(This will be derived below.) 
It can be shown that both the description of a single-nucleon via Eq.~(\ref{ustar}) and the evaluation of the 
Z-diagram, Fig.~\ref{3b}, generate a repulsive effect on the energy per particle in symmetric nuclear matter which depends on the density approximately
as 
\begin{equation}
\Delta E \propto  \left (\frac{\rho}{\rho_0}\right )^{8/3} \, , 
\label{delE} 
\end {equation}
and provides the saturating mechanism missing from conventional Brueckner calculations. 
(Alternatively, explicit TBF are used along with the BHF method in order to achieve a similar result.) 
Brown showed that the bulk of the desired effect can be obtained as a lowest order (in $p^2/m$) relativistic correction
to the single-particle propagation \citep{GB87}. 
With the in-medium spinor as in Eq.~(\ref{ustar}), the correction to the free-space spinor can be written 
approximately as 
\begin{equation}
u^*(p,\lambda) -u(p,\lambda)\approx                                                  
\left( \begin{array}{c}
 {\bf 0} \\
-\frac{\sigma \cdot \vec {p}}{2 m^2}U_S
\end{array} 
\right) \;
\chi_{\lambda},
\label{delu} 
\end{equation}
where for simplicity the spinor normalization factor has been set equal to 1, in which case it 
is clearly seen that the entire effect originates from the modification of the spinor's lower component. 
By expanding the single-particle energy to order $U_S^2$, Brown showed that the correction to the 
energy consistent with Eq.~(\ref{delu}) can be written as $\frac{p^2}{2m}(\frac{U_S}{m})^2$. He then proceeded to 
estimate the correction to the energy per particle and found it to be approximately as given in Eq.~(\ref{delE}).

The approximate equivalence of the effective-mass description of Dirac states and the contribution from the Z-diagram 
has a simple intuitive explanation in the observation 
that Eq.~(\ref{ustar}), like any other solution of the Dirac equation,
can be written as a superposition of positive and negative energy solutions. On the other hand, the "nucleon" in the 
middle of the Z-diagram, Fig.~\ref{3b}, is precisely a superposition of positive and negative energy states. 
In summary, the DBHF method effectively takes into account a particular class of 
TBF, which are crucial for nuclear matter saturation.

Having first summarized the main DBHF philosophy, 
we now proceed to describe the DBHF calculation of IANM \citep{AS03,FS10}. 
In the end, this will take us back to the crucial point of the DBHF approximation, Eq.~(\ref{ustar}). 

We start from the Thompson \citep{Thom} relativistic three-dimensional reduction 
of the Bethe-Salpeter equation \citep{BS}. The Thompson equation is applied to nuclear matter in
strict analogy to free-space scattering and reads, in the nuclear matter rest frame,                 
\begin{eqnarray}
&& g_{ij}(\vec q',\vec q,\vec P,(\epsilon ^*_{ij})_0) = v_{ij}^*(\vec q',\vec q) \nonumber \\            
&& + \int \frac{d^3K}{(2\pi)^3}v^*_{ij}(\vec q',\vec K)\frac{m^*_i m^*_j}{E^*_i E^*_j}
\frac{Q_{ij}(\vec K,\vec P)}{(\epsilon ^*_{ij})_0 -\epsilon ^*_{ij}(\vec P,\vec K)} 
g_{ij}(\vec K,\vec q,\vec P,(\epsilon^*_{ij})_0) \, ,                                   
\label{gij}
\end{eqnarray}                    
where $g_{ij}$ is the in-medium reaction matrix 
($ij$=$nn$, $pp$, or $np$), and the                                      
asterix signifies that medium effects are applied to those quantities. Thus the NN potential, 
$v_{ij}^*$, is constructed in terms of effective Dirac states (in-medium spinors) as explained above. 
In Eq.~(\ref{gij}),                                  
$\vec q$, $\vec q'$, and $\vec K$ are the initial, final, and intermediate
relative momenta, and $E^*_i = \sqrt{(m^*_i)^2 + K^2}$. 
The momenta of the two interacting particles in the nuclear matter rest frame have been expressed in terms of their
relative momentum and the center-of-mass momentum, $\vec P$, through
\begin{equation} 
\vec P = \vec k_{1} + \vec k_{2}       \label{P}    
\end{equation} 
and 
\begin{equation} 
\vec K = \frac{\vec k_{1} - \vec k_{2}}{2} \, .  \label{K}
\end{equation}                    
The energy of the two-particle system is 
\begin{equation} 
\epsilon ^*_{ij}(\vec P, \vec K) = 
e^*_{i}(\vec P, \vec K)+  
e^*_{j}(\vec P, \vec K)   
\label{eij}
\end{equation} 
 and $(\epsilon ^*_{ij})_0$ is the starting energy.
 The single-particle energy $e_i^*$ includes kinetic energy and potential 
 energy contributions.                                                           
The Pauli operator, $Q_{ij}$, prevents scattering to occupied $nn$, $pp$, or $np$ states.            
 To eliminate the angular
dependence from the kernel of Eq.~(\ref{gij}), it is customary to replace the exact
Pauli operator with its angle-average. 
Detailed expressions for the Pauli operator                     
and the average center-of-mass momentum in the case of two different Fermi seas  
can be found in Ref.~\citep{AS03}.                              

With the definitions
\begin{equation} 
G_{ij} = \frac{m^*_i}{E_i^*(\vec{q'})}g_{ij}
 \frac{m^*_j}{E_j^*(\vec{q})}             
\label{Gij}
\end{equation} 
and 
\begin{equation} 
V_{ij}^* = \frac{m^*_i}{E_i^*(\vec{q'})}v_{ij}^*
 \frac{m^*_j}{E_j^*(\vec{q})} \, ,        
\label{Vij}
\end{equation} 
 one can rewrite Eq.~(\ref{gij}) as
\begin{eqnarray}
&& G_{ij}(\vec q',\vec q,\vec P,(\epsilon ^*_{ij})_0) = V_{ij}^*(\vec q',\vec q) \nonumber \\[4pt]
&& + \int \frac{d^3K}{(2\pi)^3}V^*_{ij}(\vec q',\vec K)
\frac{Q_{ij}(\vec K,\vec P)}{(\epsilon ^*_{ij})_0 -\epsilon ^*_{ij}(\vec P,\vec K)} 
G_{ij}(\vec K,\vec q,\vec P,(\epsilon^*_{ij})_0) \, ,                                    
\label{Geq}
\end{eqnarray}                    
which is formally identical to its non-relativistic counterpart.

The goal is to determine self-consistently the nuclear matter single-particle potential   
which, in IANM, will be different for neutrons and protons. 
To facilitate the description of the procedure, we will use a schematic
notation for the neutron/proton potential.                                                   
We write, for neutrons,
\begin{equation}
U_n = U_{np} + U_{nn} \; , 
\label{un}
\end{equation}
and for protons
\begin{equation}
U_p = U_{pn} + U_{pp} \, , 
\label{up}
\end{equation}
where each of the four pieces on the right-hand-side of Eqs.~(\ref{un}-\ref{up}) signifies an integral of the appropriate 
$G$-matrix elements ($nn$, $pp$, or $np$) obtained from Eq.~(\ref{Geq}).                                           
Clearly, the two equations above are coupled through 
the $np$ component and so they must be solved simultaneously. Furthermore, 
the $G$-matrix equation and Eqs.~(\ref{un}-\ref{up})  
are coupled through the single-particle energy (which includes the single-particle
potential, itself defined in terms of the $G$-matrix). So we have a coupled system to be solved self-consistently.

Before proceeding with the self-consistency, 
one needs an {\it ansatz} for the single-particle potential. The latter is suggested by 
the most general structure of the nucleon self-energy operator consistent with 
all symmetry requirements. That is: 
\begin{equation}
{\cal U}_i({\vec p}) =  U_{S,i}(p) + \gamma_0  U_{V,i}^{0}(p) - {\vec \gamma}\cdot {\vec p}\;  U_{V,i}(p) \, , 
\label{Ui1}
\end{equation}
where $U_{S,i}$ and 
$U_{V,i}$ are an attractive scalar field and a repulsive vector field, respectively, with 
$ U_{V,i}^{0}$ the timelike component of the vector field. These fields are in general density and momentum dependent. 
We take             
\begin{equation}
{\cal U}_i({\vec p}) \approx U_{S,i}(p) + \gamma_0 U_{V,i}^{0}(p) \, ,                                            
\label{Ui2}
\end{equation}
which amounts to assuming that the spacelike component of the vector field is much smaller than 
 both $U_{S,i}$ and $U_{V,i}^0$. Furthermore, neglecting the momentum dependence of the scalar and
vector fields and inserting Eq.~(\ref{Ui2}) in the Dirac equation for neutrons/protons propagating in 
nuclear matter,
\begin{equation}
(\gamma _{\mu}p^{\mu} - m_i - {\cal U}_i({\vec p})) u_i({\vec p},\lambda) = 0  \, ,                                                       
\label{Dirac1} 
\end{equation}
naturally leads to rewriting the Dirac equation in the form 
\begin{equation}
(\gamma _{\mu}(p^{\mu})^* - m_i^*) u_i({\vec p},\lambda) = 0  \, ,                                                       
\label{Dirac2} 
\end{equation}
with positive energy solutions as in Eq.~(\ref{ustar}), $m_i^* = m + U_{S,i}$, and 
\begin{equation}
(p^0)^* = p^0 - U_{V,i}^0 (p) \, .                                                                 
\label{p0}
\end{equation}
The subscript ``$i$'' signifies that these parameters are different for protons and
neutrons. 

As in the symmetric matter case \citep{BM84}, evaluating  the expectation value of Eq.~(\ref{Ui2})       
leads to a parametrization of 
the single particle potential for protons and neutrons (Eqs.(\ref{un}-\ref{up})) in terms of the 
constants $U_{S,i}$ and $U_{V,i}^0$ which is given by      
\begin{equation}
U_i(p) = \frac{m^*_i}{E^*_i}<{\vec p}|{\cal U}_i({\vec p})|{\vec p}> = 
\frac{m^*_i}{E^*_i}U_{S,i} + U_{V,i}^0 \; .      
\label{Ui3}
\end{equation}
Also, 
\begin{equation}
U_i(p) =                                                              
\sum_{j=n,p} 
\sum_{p' \le k_F^j} G_{ij}({\vec p},{\vec p}') \; , 
\label{Ui4}
\end{equation}
which, along with Eq.~(\ref{Ui3}), allows the self-consistent determination of the single-particle
potential as explained below. 

The kinetic contribution to the single-particle energy is
\begin{equation}
T_i(p) = \frac{m^*_i}{E^*_i}<{\vec p}|{\vec \gamma} \cdot {\vec p} + m|{\vec p}> =     
\frac{m_i m^*_i + {\vec p}^2}{E^*_i} \; , 
\label{KE}    
\end{equation}
and the single-particle energy is 
\begin{equation}
e^*_i(p) = T_i(p) + U_i(p) = E^*_i + U^0_{V,i} \; . 
\label{spe}
\end{equation}
The constants $m_i^*$ and 
\begin{equation}
U_{0,i} = U_{S,i} + U_{V,i}^0      
\label{U0i} 
\end{equation}
are convenient to work with as they 
facilitate          
the connection with the usual non-relativistic framework \citep{HT70}.                       

Starting from some initial values of $m^*_i$ and $U_{0,i}$, the $G$-matrix equation is 
 solved and a first approximation for $U_{i}(p)$ is obtained by integrating the $G$-matrix 
over the appropriate Fermi sea, see Eq.~(\ref{Ui4}). This solution is 
again parametrized in terms of a new set of constants, determined by fitting the parametrized $U_i$, 
Eq.~(\ref{Ui3}), 
to its values calculated at two momenta, a procedure known as the "reference spectrum approximation". 
The iterative procedure is repeated until satisfactory convergence is reached.     

Finally, the energy per neutron or proton in nuclear matter is calculated from 
the average values of the kinetic and potential energies as 
\begin{equation}
\bar{e}_{i} = \frac{1}{A}<T_{i}> + \frac{1}{2A}<U_{i}> -m \; . 
\label{ei}
\end{equation}
 The EoS, or energy per nucleon as a function of density, is then written as
\begin{equation}
    \bar{e}(\rho_n,\rho_p) = \frac{\rho_n \bar{e}_n + \rho_p \bar{e}_p}{\rho} \, , 
\label{enp} 
\end{equation}
or 
\begin{equation}
    \bar{e}(k_F,\alpha) = \frac{(1 + \alpha) \bar{e}_n + (1-\alpha) \bar{e}_p}{2} \, . 
\label{eav} 
\end{equation}
Clearly, symmetric nuclear matter is obtained as a by-product of the calculation described above 
by setting $\alpha$=0, whereas $\alpha$=1 corresponds to pure neutron matter.

\subsubsection{Microscopic predictions of the EoS and related quantities} 

\begin{figure}
\begin{center}
\vspace*{-1.0cm}
\hspace*{-0.5cm}
\scalebox{0.3}{\includegraphics{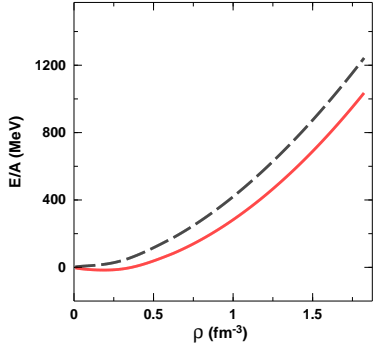}}
\vspace*{-2.5cm}
\caption{DBHF predictions for the EoS of symmetric matter (solid red) and neutron matter (dashed black).            
} 
\label{eos}
\end{center}
\end{figure}

In Fig.~\ref{eos}, we show EoS predictions for symmetric matter (solid red) and neutron matter (dashed black)  
as obtained from the Idaho calculation described in the previous section.
The EoS from DBHF can be characterized as being moderately "soft" at low to medium density                      
and  fairly "stiff" at 
high densities.                                                                                         
The predicted saturation density and energy for the symmetric matter EoS in Fig.~\ref{eos} are equal to 0.185 fm$^{-3}$ and -16.14 MeV, respectively, 
and the compression modulus is 252 MeV.                                                      

The increased stiffness featured by the DBHF EoS at the higher densities 
originates from the strongly density-dependent repulsion characteristic of the 
Dirac-Brueckner-Hartee-Fock method.  
In Ref.~\citep{Fuchs2}, it is pointed out 
that constraints from neutron star phenomenology together with flow data from heavy-ion           
reactions suggest that such EoS behavior may be desirable.

\begin{figure}
\begin{center}
\vspace*{-2.0cm}
\hspace*{-0.5cm}
\scalebox{0.4}{\includegraphics{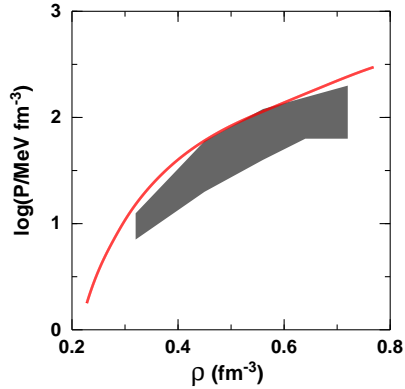}}
\vspace*{-2.5cm}
\caption{Pressure in symmetric matter from the Idaho DBHF calculation. The shaded area corresponds to the region
of pressure consistent with the flow data analysed in Ref.~\citep{MSU}.}                     
\label{psm}
\end{center}
\end{figure}

The pressure as a function of density, as discussed in the next section, plays the crucial role in building the structure of    
a neutron star. 
In Fig.~\ref{psm} we show the pressure in symmetric matter as predicted by the Idaho calculation compared with constraints obtained
from flow data \citep{MSU}. 
The predictions are seen to fall just on the high side of the constraints and grow
rather steep at high density.

\begin{figure}
\begin{center}
\vspace*{-2.0cm}
\hspace*{-0.5cm}
\scalebox{0.4}{\includegraphics{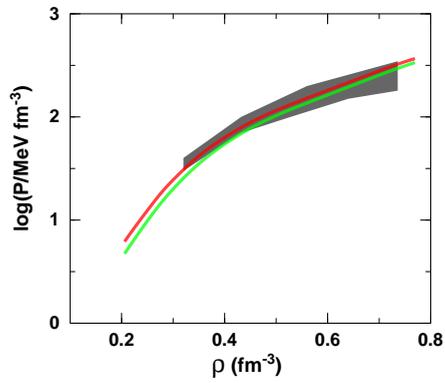}}
\vspace*{-2.5cm}
\caption{Pressure in neutron (red curve) and baryon-lepton (green curve) matter from the Idaho DBHF calculation. The
shaded area corresponds to the region
of pressure consistent with flow data and the inclusion of strong                            
density dependence in the asymmetry term \citep{MSU}.}                    
\label{pnm}
\end{center}
\end{figure}

We show in Fig.~\ref{pnm} the pressure in neutron matter (red curve)
and $\beta$-equilibrated matter (green) as predicted by DBHF calculations. The pressure contour is again from Ref.~\citep{MSU}.

\begin{figure}
\begin{center}
\vspace*{-1.0cm}
\hspace*{-1.0cm}
\scalebox{0.4}{\includegraphics{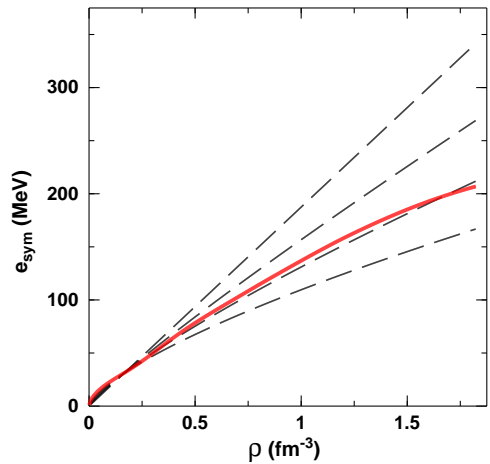}}
\vspace*{-3.0cm}
\caption{DBHF prediction for the symmetry energy (solid red) compared with various         
phenomenological parametrizations (dashed black), as explained in the text. 
} 
\label{esym}
\end{center}
\end{figure}

Next we move on to the symmetry energy as defined from Eq.~(\ref{e}). 
In Fig.~\ref{esym}, we display the Idaho DBHF prediction for the symmetry energy by the solid red curve. 
The curve is seen to grow at a lesser rate with increasing density, 
 an indication that, at large density,   
repulsion in the symmetric matter EoS increases more rapidly relative to 
the neutron matter EoS.   
This can be understood in terms of increased repulsion in isospin zero partial waves (absent
from neutron matter) as a function of density.                       
Our predicted value for the symmetry pressure $L$ (see Eq.~(\ref{L}), is close to 70 MeV.   

The various black dashed curves in Fig.~\ref{esym} are obtained with the simple parametrization               
\begin{equation}
e_{sym}=C(\rho/\rho_0)^{\gamma} \, ,  
\label{esymm} 
\end{equation}
with $\gamma$ increasing from 0.7 to 1.0 in steps of 0.1, and $C \approx 32$ MeV. 
Considering that all of the dashed curves are 
commonly used parametrizations 
suggested by heavy-ion data \citep{BA05}, 
Fig.~\ref{esym} clearly reflects our limited knowledge of the symmetry energy,      
particularly, but not exclusively, at the larger densities.                                          

As already mentioned in Section {\bf 2},      
from the experimental side intense effort is going on to obtain reliable empirical information on the less
known aspects of the EoS. Heavy-ion reactions are a popular way to seek constraints on the symmetry 
energy, through analyses of observables that are sensitive to the pressure gradient between 
nuclear and neutron matter.

\begin{figure}[!t]
\centering          
\vspace*{1.0cm}  
\includegraphics[totalheight=3.0in]{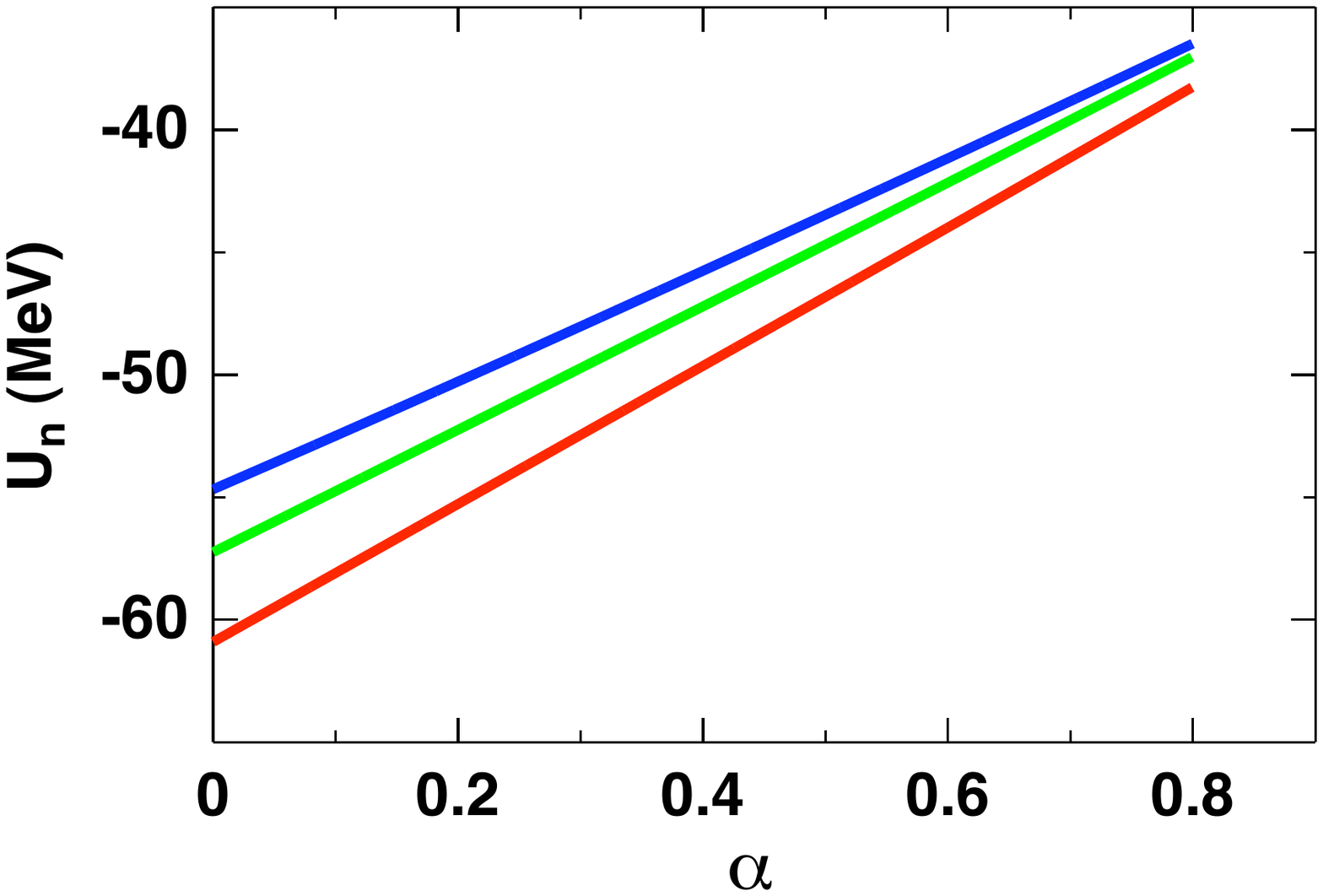}  
\includegraphics[totalheight=3.0in]{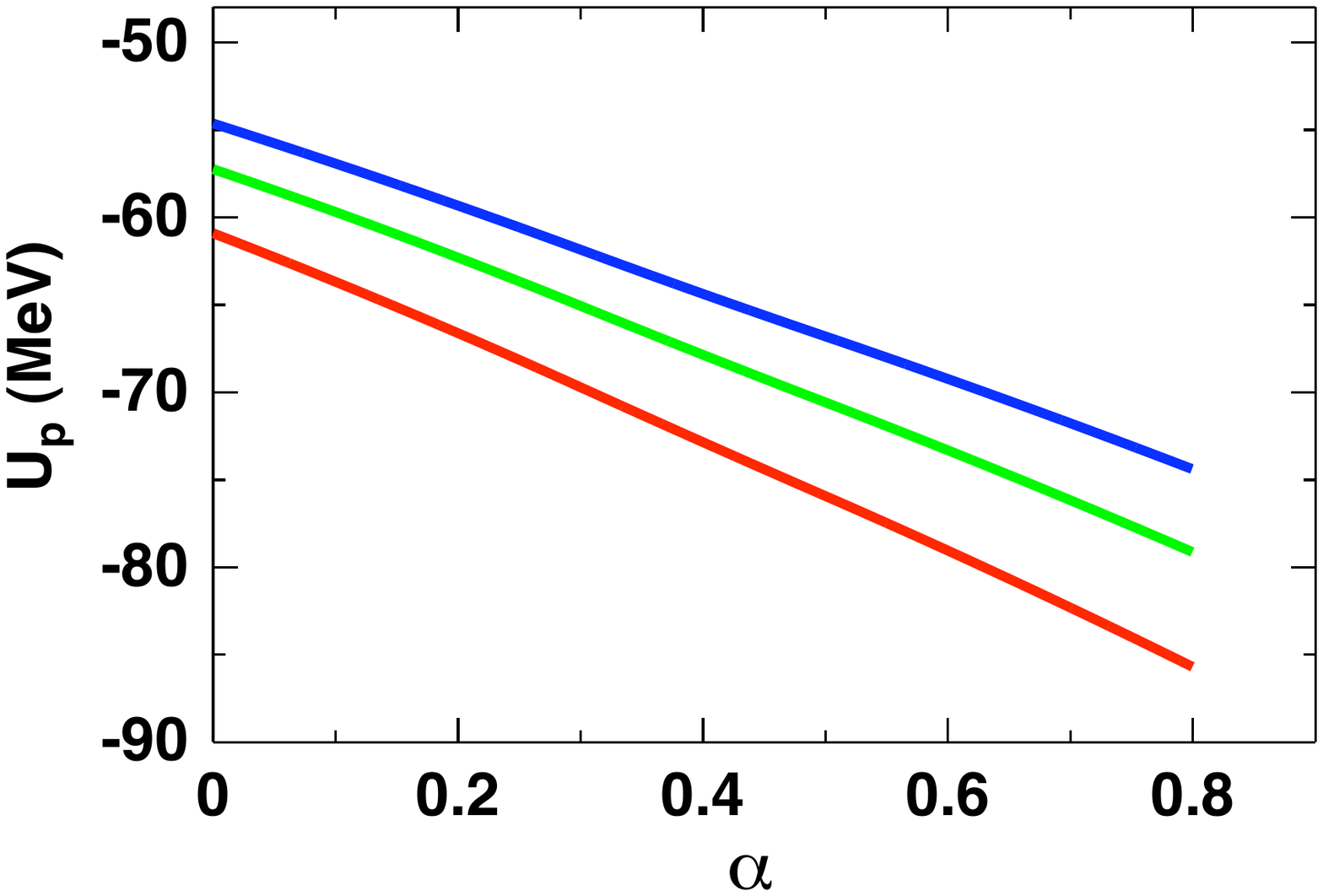}  
\vspace*{-1.5cm}
\caption{ The neutron and proton single-particle potentials as a function of the 
asymmetry parameter at fixed average density and momentum equal to the average 
Fermi momentum, which is chosen to be 1.4 fm$^{-1}$. The red, green, and blue lines
represent the predictions from the Bonn A, B, and C potentials, respectively. 
} 
\label{ualpha}
\end{figure}

We close this Section with demonstrating the approximately linear dependence 
on the asymmetry parameter
of the single-nucleon potentials in IANM as anticipated        
in Eq.~(\ref{Unp}). We recall that this isospin splitting is the crucial mechanism that separates
proton and neutron dynamics in IANM. 
In Fig.~\ref{ualpha} we display                      
predictions obtained with three different NN potentials based on the one-boson-exchange model, Bonn A, B, and C \citep{Mac89}. 
These three models differ mainly in the strength of the tensor force, which
is mostly carried by partial waves with isospin equal to 0 (absent from pure neutron matter) and thus should fade away
in the single-neutron potential
as the neutron fraction increases. In fact, the figure demonstrates reduced differences among the values               
of $U_n$ predicted with the three potentials at large $\alpha$.

\section{Neutron stars}
\subsection{A brief review of basic structure equations and available constraints}

Fusion reactions in stars give rise to elements, but at the same time exhaust the nuclear 
fuel. After the fuel is exhausted, the star can die through four possible channels:    
it can become a black hole, a white dwarf, a neutron star, or it can disassemble completely. 
The ultimate outcome depends on the mass of the original star. If the mass is larger than about
four solar masses, the star may become a supernova which, in turn, may result either in a 
neutron star or a black hole.

Neutron stars contain the most dense form of matter found in the universe and, therefore, 
are unique laboratories to study the properties of highly compressed (cold) matter. 
They are held together by
gravity and neutron degeneracy pressure. (In contrast, white dwarfs are kept 
in hydrostatic equilibrium by gravity
and electron degeneracy pressure.)                                       

Although neutron stars were predicted as early as in the 1930's, hope for their observation remained slim
for a long time. 
In 1967, strange new objects, outside the solar system, were observed at the University of Cambridge.        
They were named {\it pulsars}, as they emitted periodic radio signals. 
 To date, about 1700 pulsars have been detected, 
many in binary systems. 

Typically, 
detection of thermal radiation from the surface of a star is the way to access information about its
properties.
Furthermore, the possibility of exploring the structure of neutron stars via gravitational waves
makes these exotic objects even more interesting.

The densities found in neutron stars range from the density of iron to several times normal nuclear density. 
Most of the mass consists of 
highly compressed matter at nuclear and supernuclear densities. The {\it surface} 
region is composed of normal nuclei and non-relativistic electrons, with typical mass densities  in the range 
$10^4 \leq \epsilon \leq 10^{6}$~g~cm$^{-3}$. 
As density increases, 
charge neutrality requires matter to become more neutron rich. In this density range
(about $10^7 < \epsilon < 10^{11}$~g~cm$^{-3}$), neutron-rich nuclei appear, mostly              light metals, while electrons become relativistic. This is the {\it outer crust}.
Above densities of approximately $ 10^{11}$~g~cm$^{-3}$, free neutrons begin to form a continuum of states.
The {\it inner crust} is a compressed solid with a fluid of neutrons which drip out 
and populate free states outside the nuclei, since those have become neutron-saturated. 
Densities in 
the inner crust range between 
 $10^{11}$ and $10^{14}$~g~cm$^{-3}$.                                                    
At densities equal to approximately 1/2 of saturation density, clusters begin to merge into a continuum. In this phase, matter is a uniform fluid 
of neutrons, protons, and leptons.  
Above a few times nuclear matter density, the actual composition of stellar matter is not known.  
Strange baryons can appear when the nucleon chemical potential is of the order of their rest mass. Meson production can also take place. 
At even higher densities, transitions to other phases are speculated, such as a deconfined, rather than hadronic, phase. The critical density for
such transition cannot be predicted reliably because it lies in a range where QCD is non perturbative \citep{Sedr06}.

The possibility has been speculated that the most stable state at zero pressure may be
$u$, $d$, $s$ quark matter instead of iron. This would imply that strange quark matter is the most
stable (in fact, the absolutely stable) state of strongly interacting matter, as originally proposed 
by Bodmer \citep{Bodmer}, Witten \citep{Witten}, and Terazawa \citep{Teraz}. 
In such case, hyperonic and hybrid stars would have to be metastable with respect to stars composed 
of stable three-flavor strange quark matter \citep{Weber}, which is lower in energy 
than two-flavor quark matter due to the extra Fermi levels open to strange quarks. 
Whether or not strange quark stars can give rise to pulsar glitches (which are observed sudden small changes 
in the rotational frequency of a pulsar),
may be a decisive test of the strange quark matter hypothesis \citep{Weber}. 

The maximum gravitational mass of the star and the corresponding radius are the typical observables 
used to constraint the EoS.                                                              
The gravitational mass is inferred mostly from observations of X-ray binaries or binary
pulsars. Determination of the mass provide a unique test of both theories of nuclear
matter and general relativity.
 The pulsar in the Hulse-Taylor
binary system has a mass of 
1.4408$\pm$ 0.0003 $M_{\odot}$, to date the best mass determination.        
                   
At this time, one of the heaviest neutron stars (with        
accurately known mass) has a mass of 1.671 $\pm$ 0.008 $M_{\odot}$ \citep{Champ08}. The    
observation of an even heavier star has been
 confirmed recently, namely                  
J1614-2230, with a mass of 
1.97$\pm$0.04  $M_{\odot}$ \citep{Demorest}. This value is the highest yet measured with this certainty and represents a challenge for the softest EoS. 
We also recall that 
an initial observation of a neutron star-white dwarf binary system
had suggested a neutron star mass (PSR J0751+1807) of 2.1$\pm$0.2$M_{\odot}$ \citep{Nice}. Such observation, which 
would imply a considerable constraint on the high-density behavior of the EoS, was not confirmed. 

The minumum mass of a neutron star is also a parameter of interest. For a cold, stable system, 
the minimum mass is estimated to be 
0.09 $M_{\odot}$ \citep{Latt07}.                                       
The smallest reliably estimated neutron star mass is the companion of the binary pulsar 
J1756-2251, which has a mass of 
1.18$\pm$ 0.02 $M_{\odot}$ \citep{Faulk}.                                       

Measurements of the radius are considerably less precise than mass measurements \citep{Latt07}.
No direct measurements of the radius exist. Instead, the observed X-ray flux, together with theoretical 
assumptions \citep{Weber}, can provide information on 
the radiation or photospheric radius, $R_{\infty}$, which is related to the actual stellar radius by 
$R_{\infty} = R(1-2GM/Rc^2)^{-1/2}$.                                                  
Estimates are usually based on                                              
thermal emission of cooling stars, including redshifts, 
and the properties of sources with bursts or thermonuclear explosions at the surface. 
A major problem associated with the determination of radii is that the distance from the source is not well known, 
hence the need for additional assumptions. 
Much more stringent constraints could be imposed on the EoS if mass and radius could be 
determined independently from each other.        

Another bulk property of neutron stars is the moment of inertia, $I$. 
For softer EoS, both mass and radius are smaller and so is $I$. 
From observations of the 
Crab nebula luminosity, a lower bound on the moment of inertia was inferred to be 
$I \geq  $4-8 $\times$ 10$^{44}$ g cm$^2$, see Ref.~\citep{Weber} and references therein. 
A measurement of the moment of inertia within 10\%, together with the information on the mass, would 
be able to discriminate among various EoS \citep{Latt07}. 
To date, the best determination of the moment of inertia is the one for the Crab pulsar \citep{Crab}
which would rule out only very soft EoS \citep{Latt07}. 

A proton-neutron star is the result of a supernova explosion resulting from the gravitational collapse of a
massive star core. Nearly all of the remaining binding energy is carried away by neutrinos during the first
few tens of seconds of the evolution. 
Thus neutrino emission is very efficient as a cooling mechanism, with                                          
the internal temperature dropping to about $10^{10}$ K within a few days. 
Cooling through neutrino emission continues for a long time (in the order of 1,000 years), until the temperature
drops to about $10^{8}$ K, at which point photon emission becomes the dominant cooling mechanism.
Neutrino luminosity and emission timescale are controlled by several factors including the total mass of the (proton-neutron)
star and the opacity of neutrinos at high densities, which is sensitive to the 
EoS of dense hadronic matter.

Gravitational waves are a less conventional way to probe neutron star properties. 
Compact stars in binary systems are epected to produce gravitational radiation. In turn, 
emission of gravitational waves causes decay of the mutual orbits and eventually merger of the 
binary system. Because of the merger timescale (250 million years for PSR B1913+16, for instance,
and 85 million years for PSR J0737-3039), it can be expected that many such decaying binary systems 
exist in the galaxy and emit large amounts of gravitational radiation. The observation of 
gravitational waves has the potential to set strong constraints on masses and radii 
(see \citep{Latt07} and references therein).

A theoretical estimate of the maximum possible mass of a neutron star was performed 
by Rhoades and Ruffini \citep{RR} on the following assumptions: 1) General relativity is the 
correct theory of gravitation; 2) the EoS satisfy the Le Chatelier's principle
($\partial P/\partial \epsilon \geq 0$) and the causality condition, 
$\partial P/\partial \epsilon \leq c^2$; and 3) the EoS below some matching density 
is known. On this basis, they determined that the maximum mass of a neutron star
cannot exceed 3.2 solar masses. Abandoning the causality condition, which would hold exactly only if stellar matter is neither dispersive nor absorptive, this limit can 
be as high as 5 solar masses due to the increased stiffness of the EoS at supernuclear
densities.

The maximum mass and the radius of a neutron star are sensitive to different aspects of the EoS. The maximum mass
is mostly determined by the stiffness of the EoS at densities greater than a few times saturation density. 
The star radius is mainly sensitive to the slope of the symmetry energy. In particular, it is closely connected 
to the  internal pressure (that is, the energy gradient) of matter at
densities between about 1.5$\rho_0$ and 
2-3$\rho_0$ \citep{Latt07}.
Non-nucleonic degrees of freedom, which typically make their appearance at those densities, are
known to have a considerable impact on the maximum mass of the star. 
The latter is predicted by the equation of hydrostatic equilibrium for a perfect fluid.

In general relativity, the invariant interval between two infinitesimally close
space-time events is given by
\begin{equation}
ds^2 = g_{\alpha \beta} dx^{\alpha} dx^{\beta}                              \;, 
\label{g}
\end{equation}
where 
$ g_{\alpha \beta}$ is the space-time metric.                                  
For a spherically symmetric space-time, the most general static line 
element consistent with all required symmetries has the form
\begin{equation}
ds^2 = -f(r)dt^2 + g(r)dr^2 + h^2(r)(sin^2\theta\; d\phi^2 + d\theta^2) \;. 
\label{gr}
\end{equation}
Choosing a radial coordinate $r$ such that $h^2(r)=r^2$ yields the so-called "standard" 
 form of the metric.

The equation of hydrostatic equilibrium (the TOV equation) determines the form
of the metric functions along with the pressure and the total mass-energy density
as a function of the radial coordinate in the interior of the star. It reads
\begin{equation}
\frac{d P(r)}{dr} = -\frac{G}{c^2}\frac{(P(r)+\epsilon(r))(M(r)+4\pi r^3 P(r)/c^2)}{r(r-2GM(r)/c^2)}  \; , 
\label{GR1}
\end{equation}
with 
\begin{equation}
\frac{d M(r)}{dr} = 4\pi r^2 \epsilon(r) \, ,                           
\label{GR2}
\end{equation}
where $\epsilon$ is the total mass-energy density. 
The star {\it gravitational} mass is 
\begin{equation}
M(R) = \int_0^R 4 \pi r^2 \epsilon(r) dr \, ,                           
\label{MGR}
\end{equation}
where $R$ is the value of $r$ where the pressure vanishes. 
It's worth recalling that no mass limit exists in Newtonian gravitation.

The pressure is related to the energy per particle through 
\begin{equation}
P(\rho) = \rho ^2 \frac{\partial e(\rho)}{\partial \rho} \, .                                      
\label{Pr}
\end{equation}

The structure equations of rotationally deformed compact stars are much
more complex than those of spherically symmetric stars \citep{Weber} presented here. 
The most rapidly rotating pulsar, PSR J1748-2446 \citep{Hessels}, is believed to 
rotate at a rate of 716 Hz, 
although an X-ray burst oscillation at a frequency of 1122 Hz was reported \citep{Kaaret}, 
which may be due to the spin rate of a neutron star.
Naturally, the maximum mass and the            
(equatorial) radius become larger with increasing rotational frequency.

\subsection{Composition of $\beta$-stable matter} 
\begin{figure}[htb]	
\centering
\vspace*{-1.2cm}
\includegraphics[width=8cm]{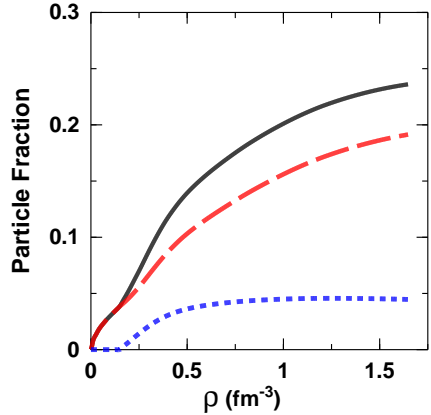} 
\vspace*{-2.2cm}
\caption{Proton (solid black), electron (dashed red), and muon (dotted blue) fractions
in $\beta$-stable matter as a function of total baryon density as predicted by the DBHF model. 
} 
\label{pfrac}
\end{figure}

Assuming that only neutrons, protons, and leptons are present, 
the proton fraction in 
stellar matter under conditions of $\beta$-equilibrium                                    
is calculated by imposing energy conservation and charge 
neutrality. The resulting algebraic equations can be found in standard literature \citep{Glen}. 
The contribution to the energy density from the electrons is written as      
\begin{equation} 
e_e= \frac{\hbar c}{4 \pi ^2}(3 \pi ^2 \rho _e)^{4/3} \, , 
\label{el}  
\end{equation}
whereas for muons we write     
\begin{equation} 
e_{\mu}= \rho _{\mu} m_{\mu}c^2 + (\hbar c)^2\frac{(3 \pi ^2 \rho_{\mu})^{5/3}}{10 \pi ^2 m_{\mu}c^2} \, .               
\label{emu}  
\end{equation}
These contributions are added to the baryonic part to give the total energy density. 
The derivative of the total energy per particle with respect to the fraction of a particular species 
is the chemical potential of that species. The conditions 
\begin{equation} 
\mu _p + \mu _e = \mu _n \, \, ; \, \,                                                                                     
\mu _{\mu} = \mu _e  \, \,  ; \, \,                                                                                      
\rho _p = \rho _{\mu} + \rho _e \; , 
\label{beta}  
\end{equation}
allow to solve for the densities (or fractions) of protons, electrons, and muons. 
Near the saturation density, when the muon fraction is close to zero, one can estimate the equilibrium 
proton fraction, $x_p$, to be \citep{Latt07} 
\begin{equation} 
x_p \approx \Big (\frac{4 e_{sym}(\rho_0)}{\hbar c} \Big )^3/(3 \pi ^2 \rho _0). 
\label{xp}  
\end{equation}
The fractions of protons, electrons, and muons as predicted with the  DBHF equation of state are shown in Fig.~\ref{pfrac}. 
The critical density for the proton fraction to exceed approximately $1/9$ and, thus, allow cooling through the  direct Urca  processes,                          
\begin{equation} 
n \rightarrow p + e + \bar{\nu}_e \, \, \, \, \, and \, \, \, \, \, 
 p +e \rightarrow n + {\nu}_e \, ,                           
\label{Urca}  
\end{equation}
is about 
$0.36-0.39$ fm$^{-3}$. 
Notice that, due to the relation between symmetry energy and proton fraction,
large values of the symmetry energy would make the star cool rapidly. 
In fact, already in earlier studies \citep{Boguta} tha rapid cooling of neutron stars 
and the corresponding high neutrino luminosity was understood in terms of neutron 
$\beta$ decay and large proton fractions. 
                                                                     
At densities close to normal nuclear density, protons and neutrons are the only baryonic degrees
of freedom. As density increases, other baryons begin to appear, such as strange baryons 
or isospin 3/2 nucleon resonances. 
 Hyperonic states can be classified according to the irreducible representation of the 
$SU(3)$ group. The octet of baryons that can appear in neutron matter includes nucleons,
$\Lambda$, $\Sigma ^{0,\pm}$, and $\Xi ^{0,-}$. 

Neglecting the nucleon-hyperon interaction, the threshold for stable hyperons to exist in matter is       
determined by comparing the hyperon mass with the neutron Fermi energy, which is the largest available energy scale 
in neutron-rich matter.                                                                                         
We consider cold neutron stars, after neutrinos have escaped. Strange baryons appear at about 2-3 times
normal density \citep{nycatania2}, an estimate which is essentially model independent, through the
processes $n + n \rightarrow p + \Sigma ^-$ and 
 $n + n \rightarrow n + \Lambda$.      
The equilibrium conditions for these reactions are
\begin{equation} 
2 \mu _n = \mu _p + \mu _{\Sigma^-} \, \, ; 
\, \, \mu _n = \mu _{\Lambda}\, .                     
\label{nybeta1} 
\end{equation} 
Also, we have 
\begin{equation} 
 \mu _e = \mu _{\mu} \, \, ; 
\, \, \mu _n = \mu _p + \mu _e \, ,                     
\label{nybeta1b} 
\end{equation} 
the equations above being special cases of 
\begin{equation} 
 \mu  = b\mu _{n} - q \mu _e \, , 
\label{nybeta1c} 
\end{equation} 
where $b$ and $q$ are the baryon number and the charge (in units of the electron charge)
of the particular species with chemical potential $\mu$. 
Together with the charge neutrality condition and baryon number conservation,
\begin{equation} 
\rho _p = \rho _e + \rho _{\mu} + \rho _{\Sigma ^-} \, \, ; 
\, \, \rho  = \rho _n + \rho _p + \rho _{\Sigma ^-}              
+ \rho _{\Lambda}\, ,        
\label{nybeta2} 
\end{equation} 
the above system allows to determine the various particle fractions.

Naturally, 
the composition of matter at supra-nuclear densities determines the behavior of stellar matter.
It is also speculated that a transition to a quark phase may take place at very high densities, 
the occurrence of which depends sensitively on the properties of the EoS in the hadronic (confined)
phase.                 
The presence of hyperons in the interior of neutron stars is reported to 
soften the equation of state, with the consequence that the predicted             
neutron star maximum masses become considerably smaller \citep{Sch+06}.            
Strange baryons are not included in the predictions shown below.

\subsection{Microscopic predictions of neutron star properties}
We are now ready to move to applications of our EoS to compact stars.                

\begin{figure}[!t] 
\centering          
\includegraphics[totalheight=5.8in]{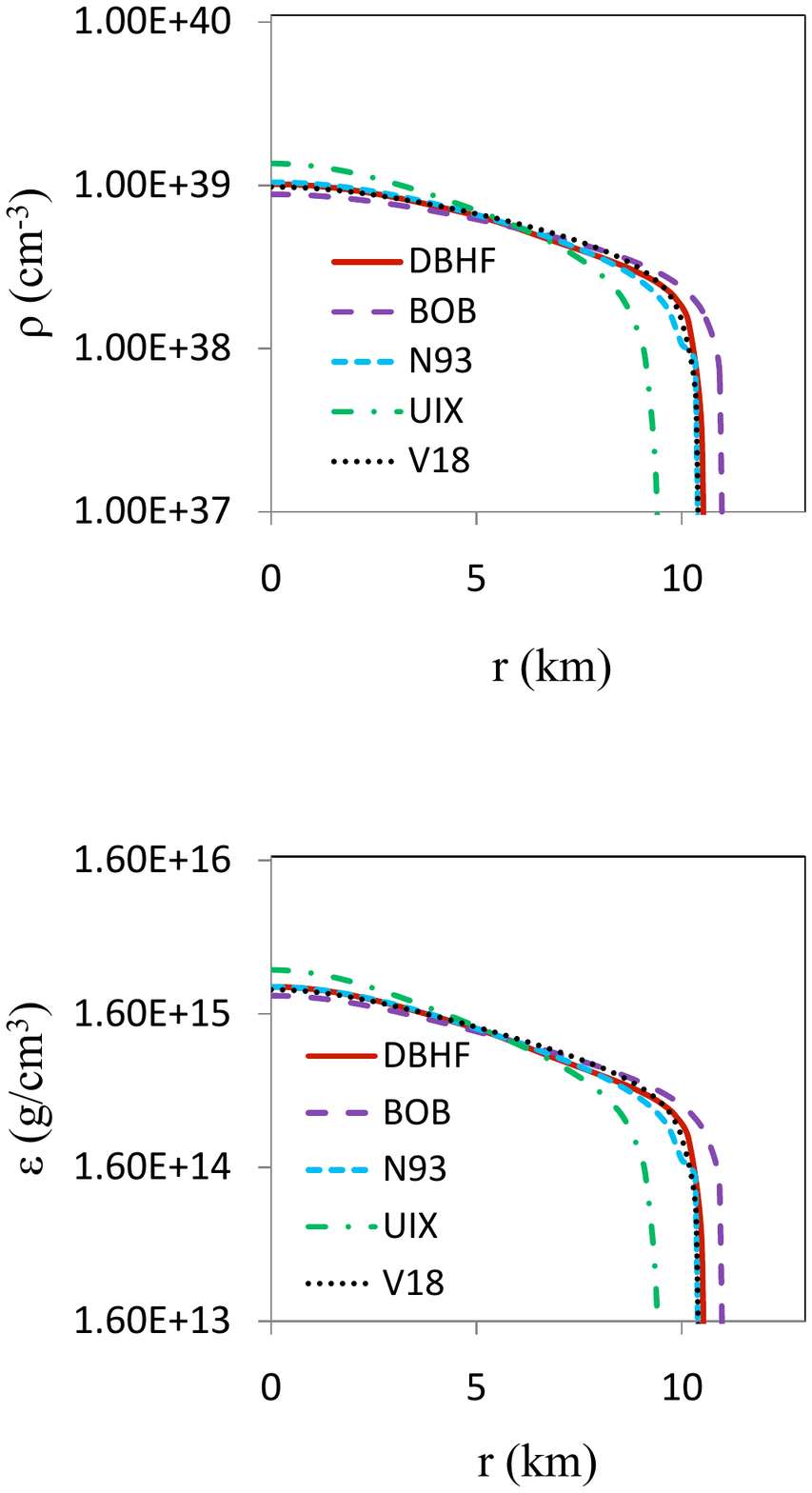} 
\vspace*{-3.2cm}
\caption{The baryon density and the mass-energy density profile for a neutron star     
with the maximum mass allowed by each EoS model. 
} 
\label{rprof}
\end{figure}

\begin{figure}[!t] 
\centering          
\includegraphics[totalheight=5.8in]{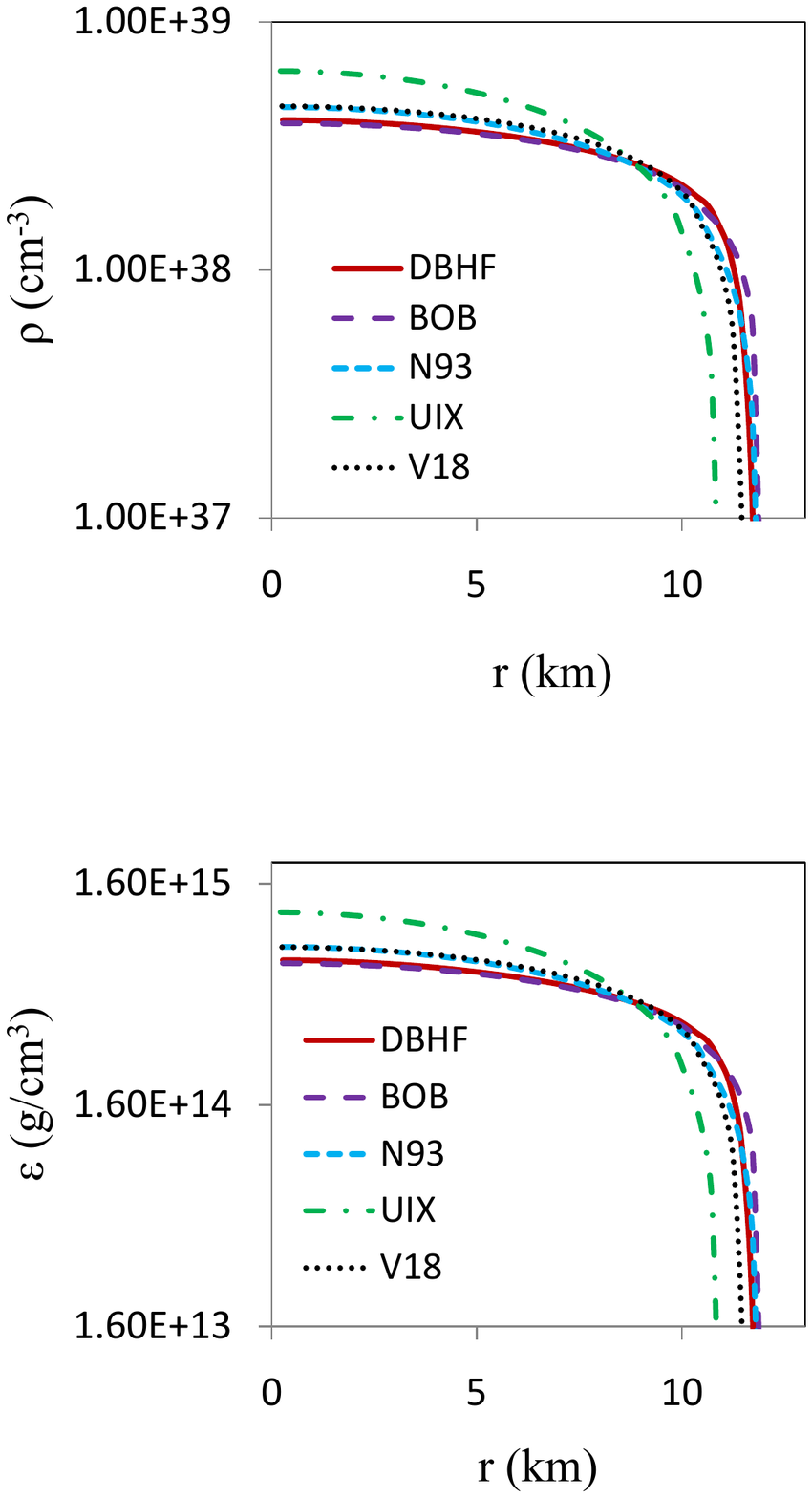} 
\vspace*{-3.2cm}
\caption{The baryon density and the mass-energy density profile for a neutron star     
with a mass of 1.4 solar masses.                 
} 
\label{rprof1.4}
\end{figure}

\begin{figure}[!t] 
\centering          
\includegraphics[totalheight=6.8in]{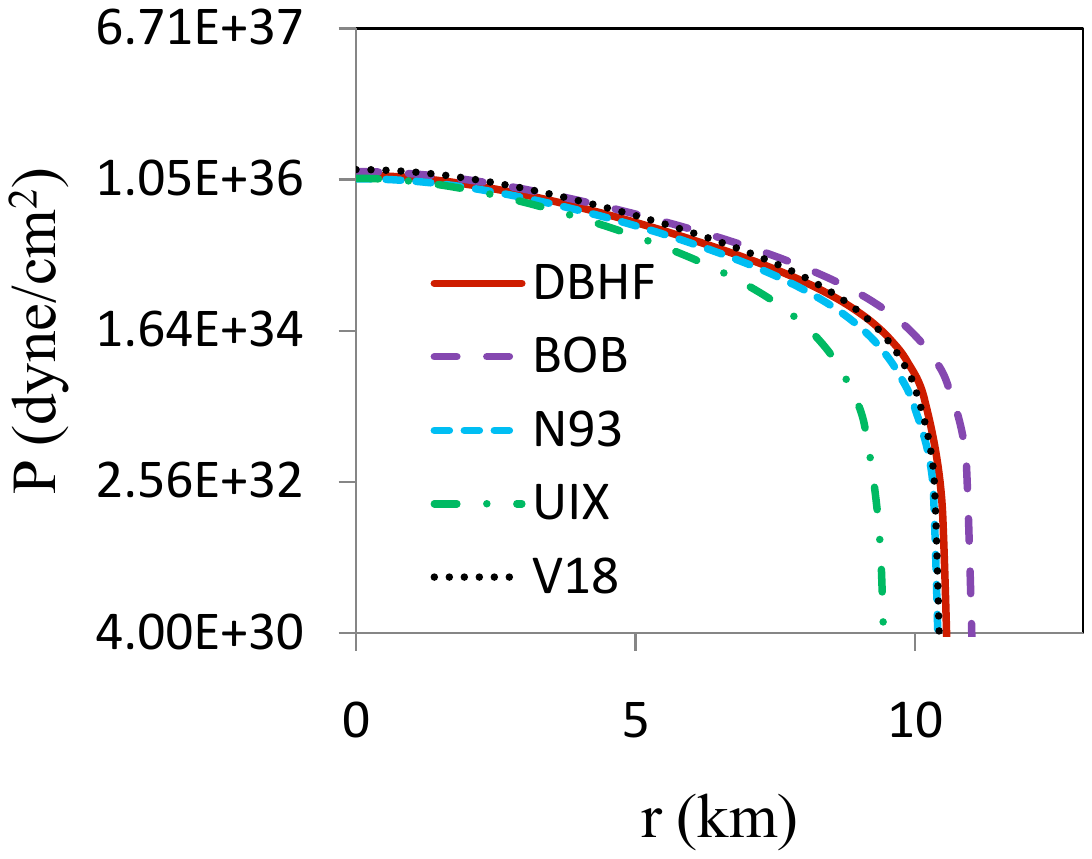} 
\vspace*{-10.2cm}
\caption{Pressure profile for the maximum-mass star allowed by each EoS model. 
} 
\label{prprof}
\end{figure}

\begin{figure}[!t] 
\centering          
\includegraphics[totalheight=6.8in]{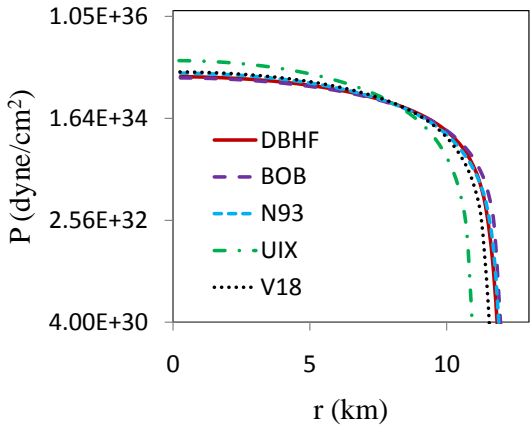} 
\vspace*{-10.2cm}
\caption{Pressure profile predicted by the various models for a 1.4 solar mass star.                  
} 
\label{prprof14}
\end{figure}

\begin{figure}[!t] 
\centering          
\includegraphics[totalheight=5.8in]{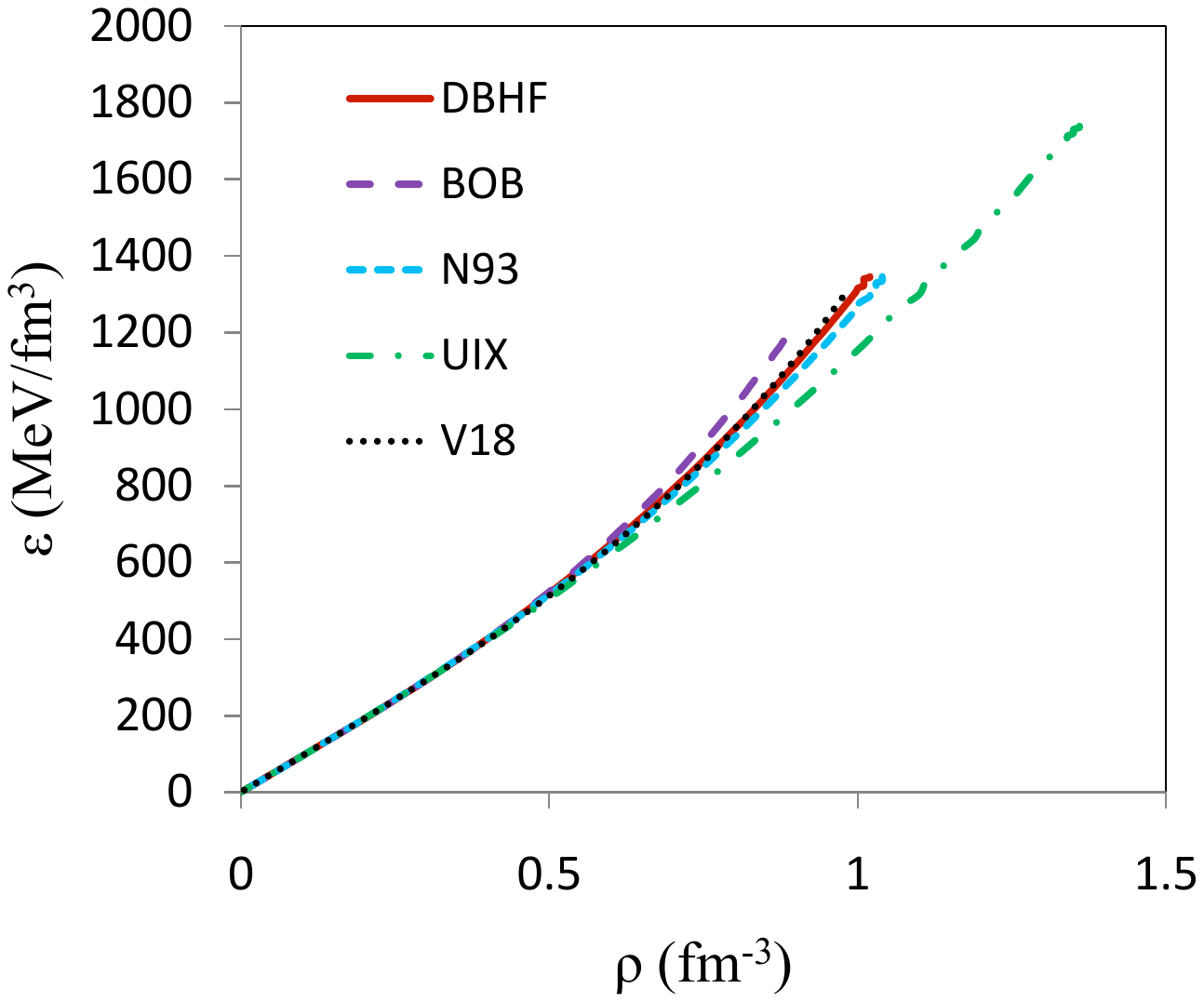} 
\vspace*{-8.0cm}
\caption{Energy density {\it vs.} baryon number density for the various EoS
 being addressed in the text. The maximum-mass model is considered in each case. 
} 
\label{edrho}
\end{figure}

As explained in Section {\bf 3},       
the DBHF model does not include three-body forces explicitely, but effectively incorporates the 
class of TBF originating from the presence of nucleons and antinucleons (the          
"Z-diagrams" in Fig.~\ref{3b}),               
see discussion in Section {\bf 3.1}. 
In order to broaden our scopes, we will compare our predictions with those of 
other microscopic models. 
As the other element of our comparison, we will take the EoS from the microscopic approach 
of Ref.~\citep{Catania3}. There (and in previous work by the same authors), 
the Brueckner-Hartree-Fock (BHF) formalism is employed along with microscopic three-body forces.        
In particular, in Ref.\citep{Catania4}
the meson-exchange TBF are constructed applying the same parameters
as used in the corresponding nucleon-nucleon potentials, which are: Argonne V18 \citep{V18} (V18), Bonn B \citep{Mac89} (BOB), Nijmegen 93 \citep{N93} (N93). 
The popular (but phenomenological) Urbana TBF \citep{UIX} (UIX) is also utilized in Ref.~\citep{Catania4}. Convenient
parametrizations in terms of simple analytic functions are given in all cases and we
will use those to generate the various EoS. We will refer to this approach, generally, as "BHF + TBF". 
                                  
At subnuclear densities all the EoS considered here are joined with the crustal equations of state from Harrison and Wheeler \citep{HW} and Negele and Vautherin \citep{NV}. 
The composition of the crust is crystalline, with light \citep{HW} or heavy \citep{NV} metals and electron gas.

We begin with showing the baryon number density and the mass-energy density profile 
of the star, see Fig.~\ref{rprof}. For each EoS model, the maximum mass configuration is considered. Thus the models differ in their central density, which, in turn, impact the 
radius. 
The relations shown in Fig.~\ref{rprof} are insightful, 
as they reveal the detailed structure of the star at each radial position.
Furthermore, the compactness of the star, whose density profile is reminescent of the 
one in a nucleus, a system 55 orders of magnitude lighter, is apparent. 
In Fig.~\ref{rprof1.4}, the same quantities are shown for a star with a mass of 1.4 solar masses, the most      
probable mass of a neutron star.                       

The models labeled as UIX and BOB have the smallest and largest radius, respectively, as
can be seen from the figure. We also see that the star's outer regions, that is, for energy densities less than about 10$^{14}$g cm$^{-3}$,  are influenced quite
strongly by differences in the various EoS models. 
Note that the UIX model, with the smallest radius, can tolerate larger central 
densities.

Of interest is also the pressure profile for the maximum-mass star in each model, which  
is shown in Fig.~\ref{prprof} for the maximum mass and in Fig.~\ref{prprof14} for a 1.4 solar mass star.    

The $\epsilon(r)$ and $\rho(r)$ relations from Fig.~\ref{rprof} are combined to provide the 
$\epsilon(\rho)$ relation within the star as shown in Fig.~\ref{edrho} for the                          
maximum mass.                                         
Again, we see that the stiffest (BOB) and softest EoS (UIX)                          
support the smallest and largest central densities, respectively. At the same time,
these two EoS predict the largest (BOB) and smallest (UIX) maximum mass, see below.  
            
In Fig.~\ref{MR}, we show the mass-radius relation for a sequence of static neutron stars as predicted
by the various models.                                                            
All models besides DBHF share the same many-body
approach (BHF+TBF) but differ in the two-body potential and TBF employed. The differences resulting from the 
use of different NN potentials can be 
larger than those originating from emplying different many-body approaches. This can be seen by comparing
the DBHF and BOB curves, both employing the Bonn B interaction (although in the latter case the non-reltivistic, r-space version of the potential is adopted). Overall, the maximum masses range from 1.8$M_{\odot}$
(UIX) to 
2.5$M_{\odot}$ (BOB). Radii are less sensitive to the EoS and range between 10 and 12 km for all models under
consideration, DBHF or BHF+TBF. 
Concerning consistency with present constraints, the observations reported in Section {\bf 4.1} 
would appear 
to invalidate only the model with the smallest maximum mass, UIX. 
Notice, further, that phenomena such as condensation of mesons may soften the EoS  
considerably at supernuclear density as condensation would bring loss of pressure. 

\begin{figure}[!t] 
\centering          
\includegraphics[totalheight=4.5in]{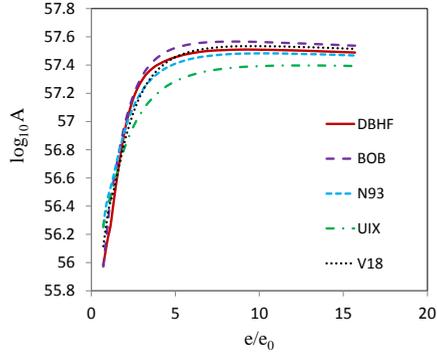} 
\vspace*{-5.3cm}
\caption{Baryon number as a function of the central density (in units of   
$e_0=2.5\;10^{14}$ g cm$^{-3}$) for the models 
considered in the text. 
} 
\label{Abar}
\end{figure}

\begin{figure}[!t] 
\centering          
\includegraphics[totalheight=4.5in]{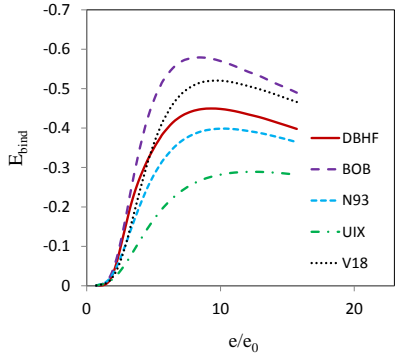} 
\vspace*{-5.3cm}
\caption{Binding energy (in units of the solar mass) as a function of the central density for the EoS models
considered in the text.                                                               
} 
\label{Eb}
\end{figure}

Also of interest is the star baryon number, $A$, which is obtained by integrating the 
baryon density over the proper volume \citep{Weber}. Namely, 
\begin{equation}
A = 4 \pi \int_0^R dr \;r^2 \frac{\rho(r)}{\sqrt{1 - 2 GM(r)/(rc^2)}} \;. 
\end{equation} 
Defining the star's baryon mass as
\begin{equation}
M_A = m_n\,A \;,
\label{mb}
\end{equation} 
where $m_n$ is the mass of the baryon, 
one calculates the star's binding energy as 
\begin{equation}
E_B = M - M_A \; .                      
\label{EB} 
\end{equation} 
The baryon number and the star binding energy as a function of the central density (in units of nuclear matter density) are shown in 
 Fig.~\ref{Abar} and Fig.~\ref{Eb} for the various models. 
We see that the baryon number for stable stars is approximately equal to $10^{56}-10^{57}$. 
A much higher value would make the star unstable with respect to gravitational collapse.
 The binding energy displayed in Fig.~\ref{Eb} is defined in units of the solar mass. From the 
definition, Eq.~\ref{EB}, negative values signify a bound system. Typically, the binding energy changes the sign for masses
less than 0.1 solar masses. 
The binding energy is a potentially observable quantity, since 
neutrinos from a supernova            
carry information about the gravitational binding energy of the neutron star that has resulted from the explosion.

\begin{figure}[!t] 
\centering          
\includegraphics[totalheight=5.5in]{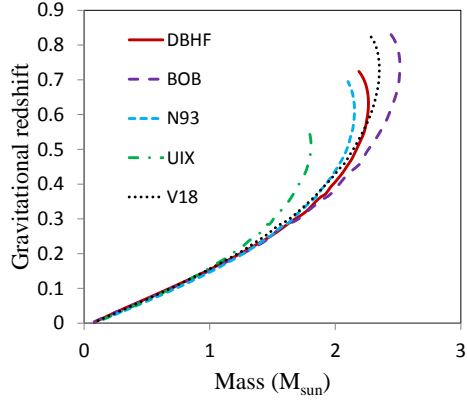}                
\vspace*{-7.1cm}
\caption{ Gravitational redshift for all models. 
For each model, the corresponding sequence of static stars is considered. 
} 
\label{Z}
\end{figure}

Next, we calculate the gravitational redshift predicted by each model. 
The redshift is defined as     
\begin{equation}
z = \frac{\nu_E}{\nu_{\infty}} -1                 \; , 
\label{red1}
\end{equation}
where $\nu_e$ and 
$\nu_{\infty}$ are the photon frequencies at the emitter and at the infinitely far
receiver. 
The photon frequency at the emitter  is 
the inverse of the proper time between two wave crests in the frame of the emitter,           
\begin{equation}
\frac{1}{d\tau_E} = (-g_{\mu \nu}dx^{\mu}dx^{\nu})_E^{-1/2} \;, 
\label{red2}
\end{equation}
with a similar expression for the frequency at the receiver. 
Then 
\begin{equation}
\frac{\nu_{\infty}}{\nu_E} = \frac{((-g_{00})^{1/2}dx^0)_E} 
{((-g_{00})^{1/2}dx^0)_{\infty}}\; . 
\label{red3}
\end{equation}
Assuming a static gravitational field, in which case the time $dx^0$ between two crests is the same at the star's surface and at the receiver, and writing  $g_{00}$ as the metric tensor component 
at the surface of a nonrotating star yield the simple equation 
\begin{equation}
z = \Big (1 - \frac{2MG}{Rc^2} \Big ) ^{-1/2} -1 \; . 
\label{red}
\end{equation}
Notice that simultaneous measurements of $R_{\infty}$ and $z$ 
determines both $R$ and $M$, since
\begin{equation}
R = R_{\infty}(1+z)^{-1}                         \; , 
\label{red4}
\end{equation}
and 
\begin{equation}
M = \frac{c^2}{2G}R_{\infty}(1+z)^{-1}[1-(1+z)^{-2}] \; . 
\label{red5}
\end{equation}
In Fig.~\ref{Z} we show the gravitational redshift as a function the mass for each model.               
Naturally the rotation of the star modifies the metric, and in that case different considerations need to
be applied which result in a frequency dependence of the redshift. We will not consider the general case
here.

\begin{figure}[!t] 
\centering          
\includegraphics[totalheight=2.2in]{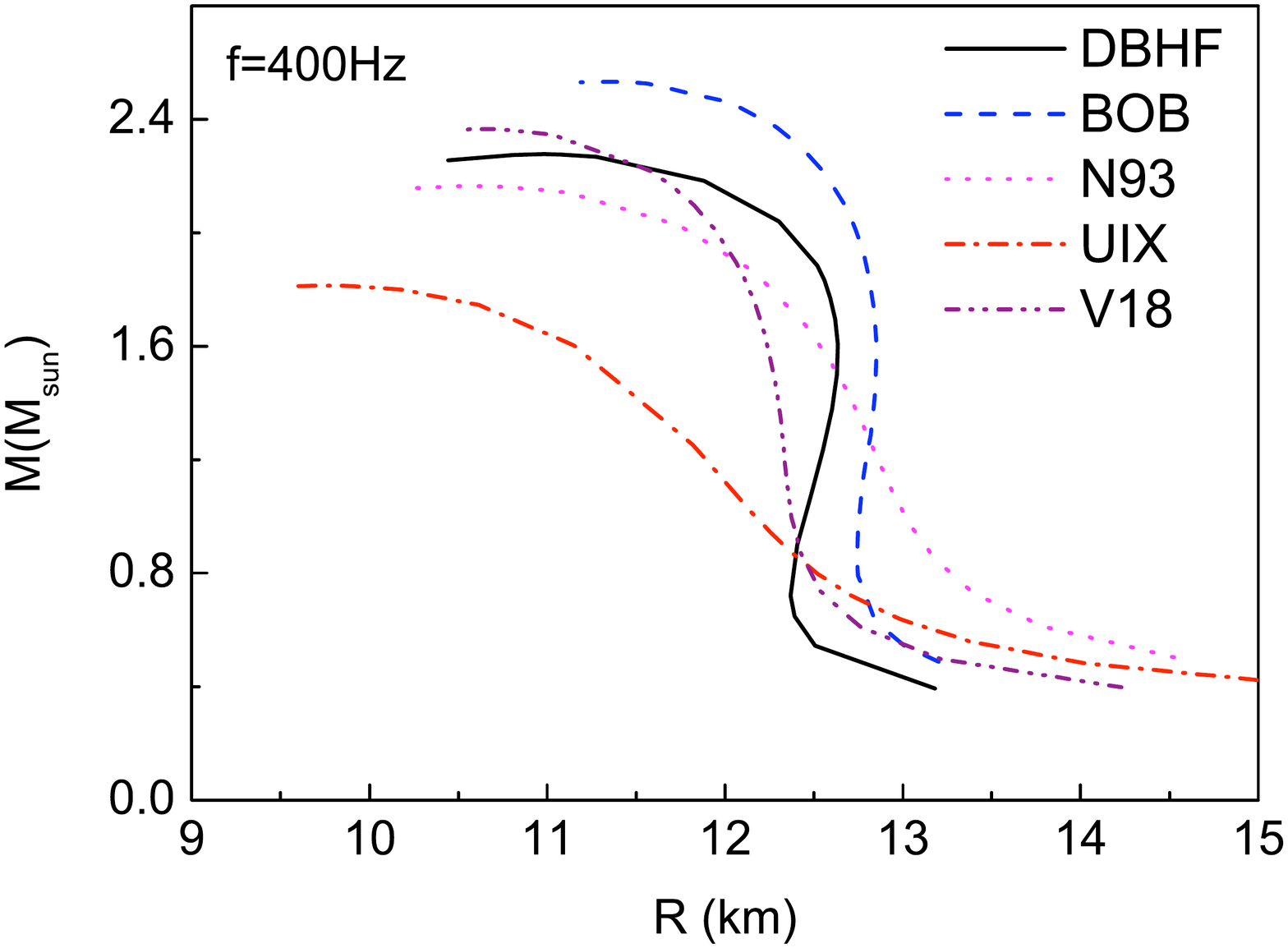}  
\includegraphics[totalheight=2.2in]{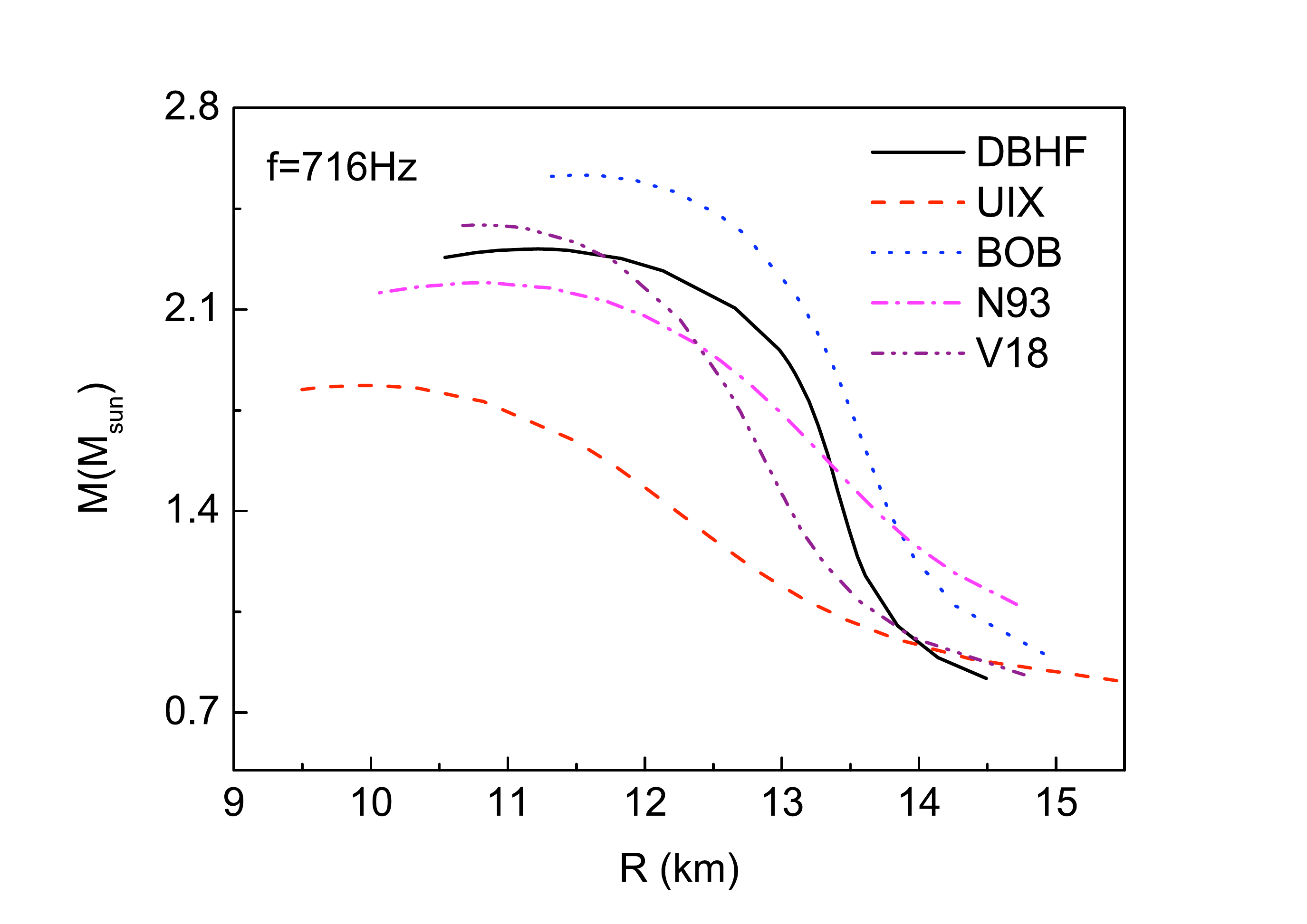}  
\includegraphics[totalheight=2.2in]{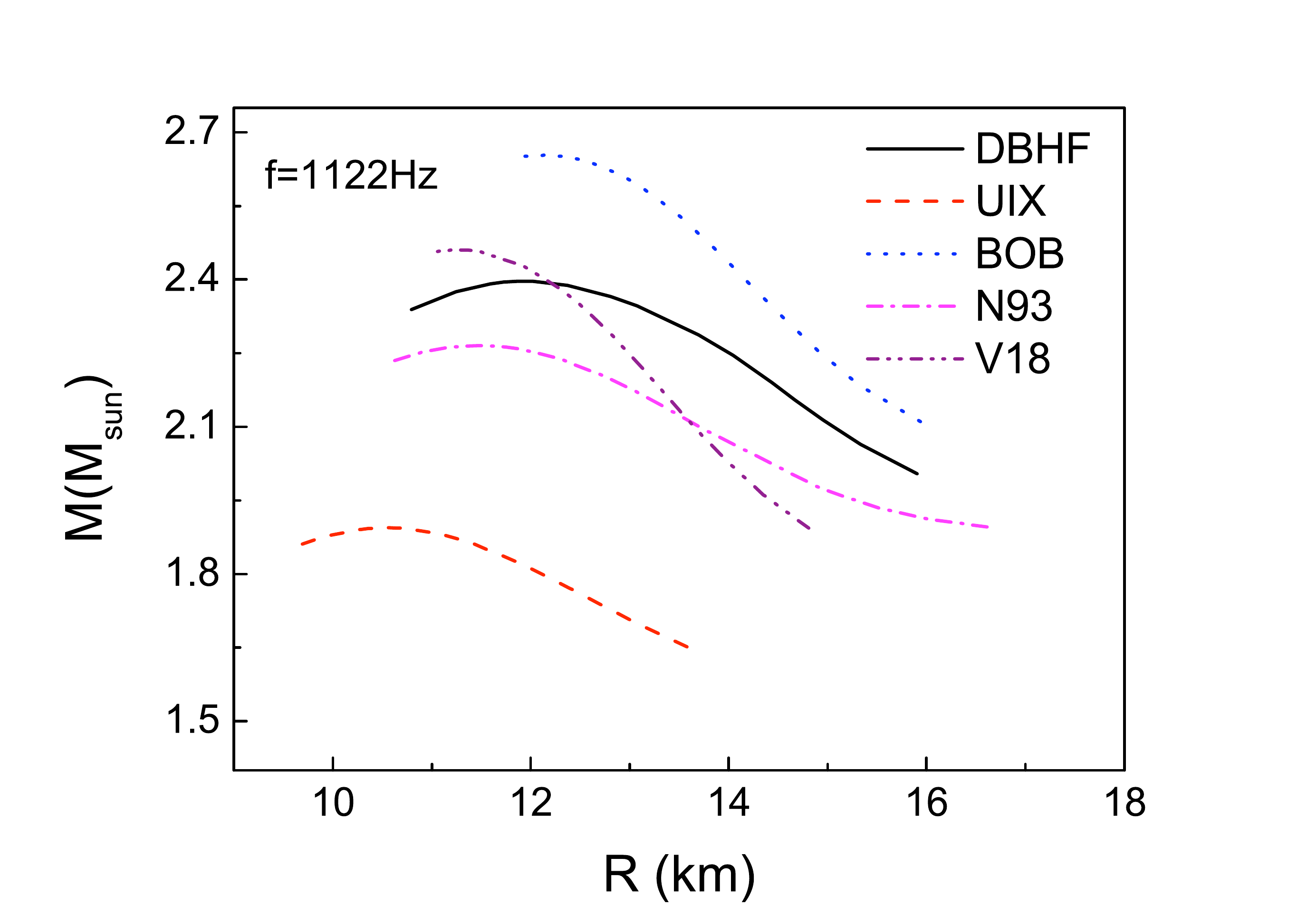}    
\vspace*{0.2cm}
\caption{Mass-radius relation for the models
considered in the text and for different rotational frequencies.                  
} 
\label{MRrot}
\end{figure}

We conclude this section with showing 
a few predictions 
for the case of rapidly rotating stars. 
The model dependence of the mass-radius relation is shown in Fig.~\ref{MRrot}.                
The 716 Hz frequency corresponds to the most rapidly rotating pulsar, PSR J1748-2446,\citep{Hessels}  
although recently an X-ray burst oscillation at a frequency of 1122 Hz has been reported\citep{Kaaret}
which may be due to the spin rate of a neutron star.
As expected, the maximum mass and the            
(equatorial) radius become larger with increasing rotational frequency.

\begin{figure}[!t] 
\centering          
\includegraphics[totalheight=2.2in]{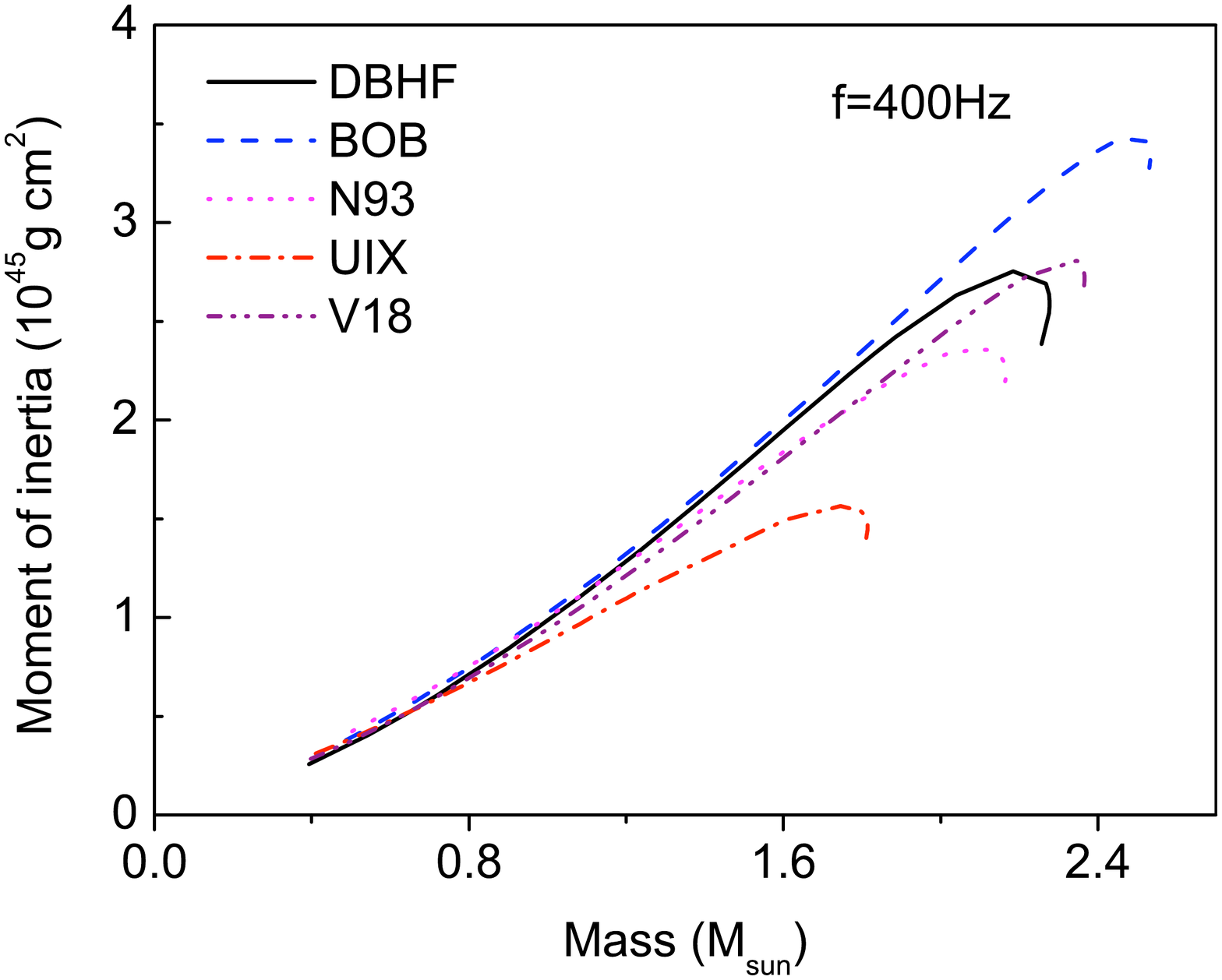}  
\includegraphics[totalheight=2.2in]{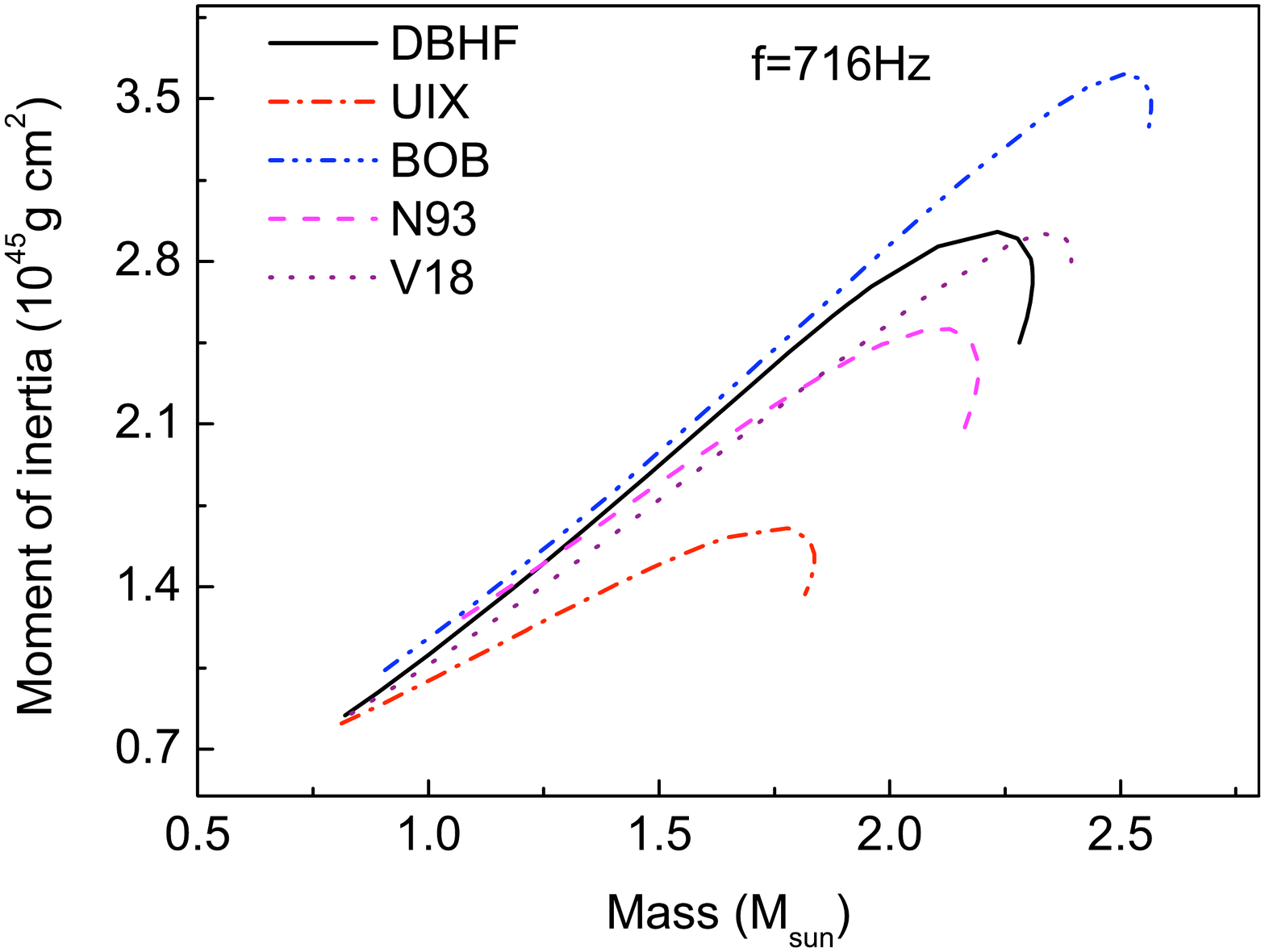}  
\includegraphics[totalheight=2.2in]{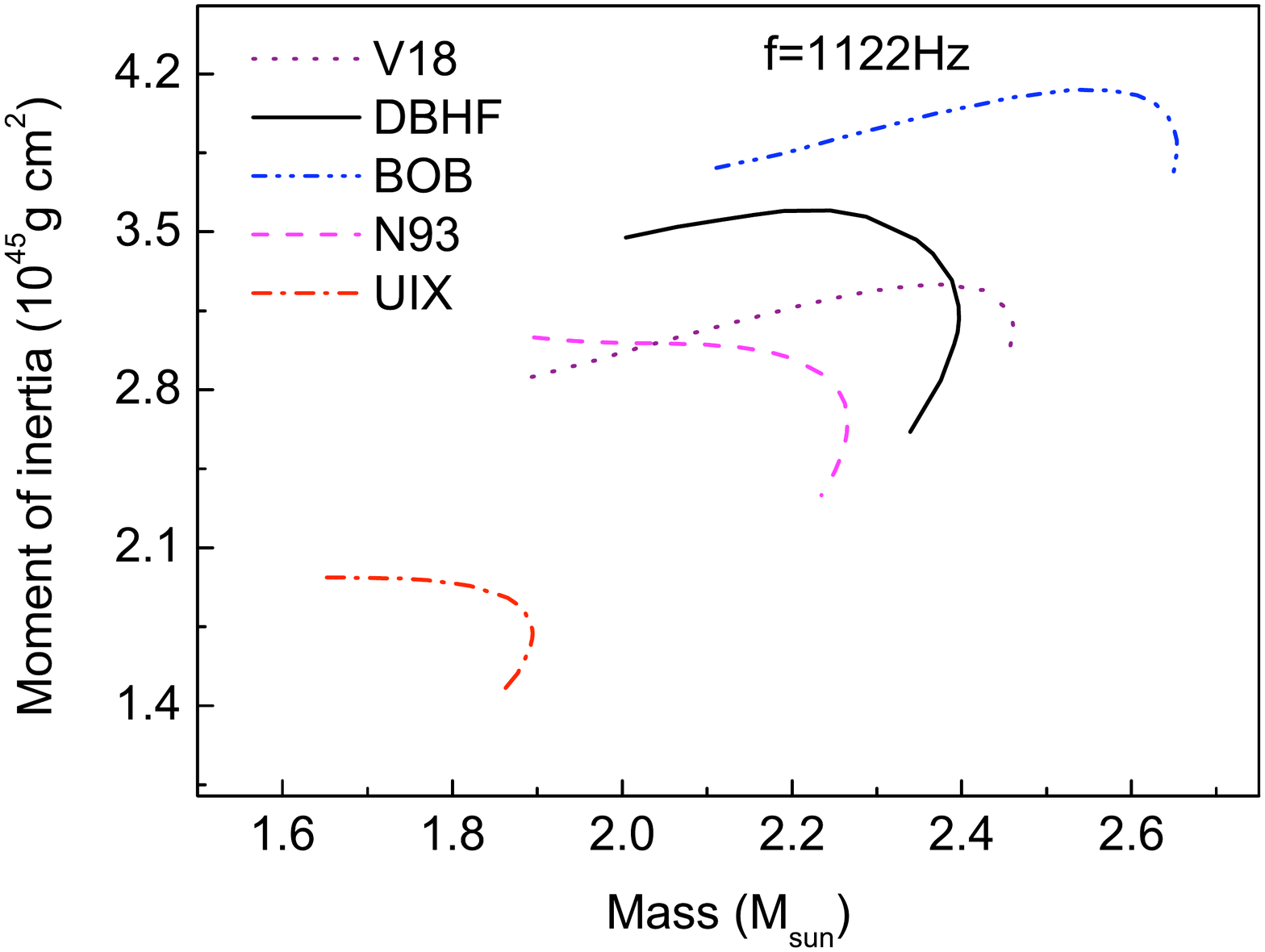}   
\vspace*{0.2cm}
\caption{ Moment of inertia for  the models
considered in the text and for different rotational frequencies.                  
} 
\label{IM}
\end{figure}

In Fig.~\ref{IM}, we show the moment of inertia at different rotational speeds (again, for all models).                 
These values are not in contradiction with observations of the 
Crab nebula luminosity, from which a lower bound on the moment of inertia was inferred to be 
$I \geq  $4-8 $\times$ 10$^{44}$ g cm$^2$, see Ref.\citep{Weber} and references therein.

Clearly, at the densities           
probed by neutron stars the model dependence is large, but presently available constraints are 
still insufficient to discriminate among these EoS. 
The model dependence we observe comes from two sources, the two-body potential and the 
many-body approach, specifically the presence of explicit TBF or Dirac effects. 
The dependence on the two-body potential is very large. Typically, the main source of model dependence 
among NN potentials is found in the strength of the tensor force.                                            
Of course, differences at the two-body level impact the TBF as well, whether they are microscopic
or phenomenological.

\begin{figure}[!t] 
\centering          
\includegraphics[totalheight=5.5in]{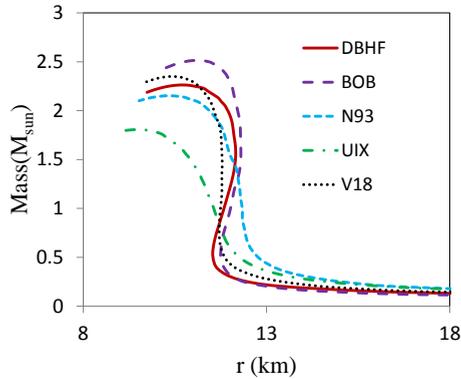} 
\vspace*{-7.2cm}
\caption{Static neutron star mass-radius relation for the models
considered in the text. 
} 
\label{MR}
\end{figure}

\section{Polarized isospin-asymmetric matter} 
Before concluding this chapter, we like to touch upon the issue of 
polarization in IANM. 
                                              
When both isospin and spin asymmetries are present,                                          
constraints are much more difficult to obtain and predictions regarding magnetic properties of nuclear matter 
are sometimes found to be in qualitative disagreement with one another. This is especially the case 
with regard to 
the possibility of spontaneous phase transitions into spin ordered states, ferromagnetic (FM, 
with neutron and proton spins alligned), 
or antiferromagnetic state (AFM, with opposite spins for neutrons and protons). 
Notice that 
the presence of polarization would impact neutrino cross section and luminosity, resulting into a very 
different scenario for neutron star cooling, which is why we find it appropriate to briefly discuss
this issue here. 

Recently, 
we have extended the framework described in Section {\bf 3} \citep{FS11b} 
to include both spin and isospin asymmetries of nuclear
matter and calculated the energy per particle 
under extreme conditions of polarization.                  
The existence (or not) of a possible phase transition can be argued by comparing the 
energies of the fully polarized and the unpolarized phases. 

In a spin-polarized and isospin asymmetric system with fixed total density, $\rho$,           
the partial densities of each species are               
\begin{equation}
\rho_n=\rho_{nu}+\rho_{nd}\; , \; \; \; 
\rho_p=\rho_{pu}+\rho_{pd}\;, \; \; \; 
\rho=\rho_{n}+\rho_{p} \; ,           
\end{equation}
where $u$ and $d$ refer to up and down spin-polarizations, respectively, of protons ($p$) or neutrons ($n$). 
The isospin and spin asymmetries, $\alpha$, $\beta_n$, and $\beta_p$,  are defined in a natural way: 
\begin{equation}
\alpha=\frac{\rho_{n}-\rho_{p}}{\rho} \;, \; \; \;
\beta_n=\frac{\rho_{nu}-\rho_{nd}}{\rho_n} \;, \; \; \; 
\beta_p=\frac{\rho_{pu}-\rho_{pd}}{\rho_p} \;. 
\end{equation}

The single-particle potential of a nucleon in a particular $\tau \sigma$ state, $U_{\tau \sigma}$, is now the solution of a
set of four coupled equations, which are the appropriate extension of Eqs.~(\ref{un}-\ref{up}). 
They read 
\begin{equation}
U_{nu} = U_{nu,nu} + U_{nu,nd} + U_{nu,pu} + U_{nu,pd}       
\end{equation} 
\begin{equation} 
U_{nd} = U_{nd,nu} + U_{nd,nd} + U_{nd,pu} + U_{nd,pd}        
\end{equation} 
\begin{equation} 
U_{pu} = U_{pu,nu} + U_{pu,nd} + U_{pu,pu} + U_{pu,pd}      
\end{equation} 
\begin{equation} 
U_{pd} = U_{pd,nu} + U_{pd,nd} + U_{pd,pu} + U_{pd,pd}   \; , 
\end{equation}
to be solved self-consistently along with the two-nucleon $G$-matrix.                   
In the above equations,                                                              
each $U_{\tau \sigma, \tau '\sigma'}$ term contains the
appropriate (spin and isospin dependent) part of the interaction, $G_{\tau \sigma,
'\tau'\sigma'}$. More specifically,
\begin{equation}
U_{\tau \sigma}({\vec k}) = \sum _{\sigma '=u,d}\sum_{\tau'=n,p} \sum _{q\leq k_F^{\tau' \sigma
'}} <\tau \sigma,\tau'\sigma'|G({\vec k},{\vec q})|\tau \sigma,\tau'\sigma'>,
\end{equation}
where the third summation indicates integration over the Fermi
seas of protons and neutrons with spin-up and spin-down. Notice that this equation is the 
extension of Eq.~(\ref{Ui4}) in the presence of spin polarization.

\begin{figure}[!t] 
\centering 
\vspace*{-1.0cm}
\hspace*{-1.0cm}
\scalebox{0.25}{\includegraphics{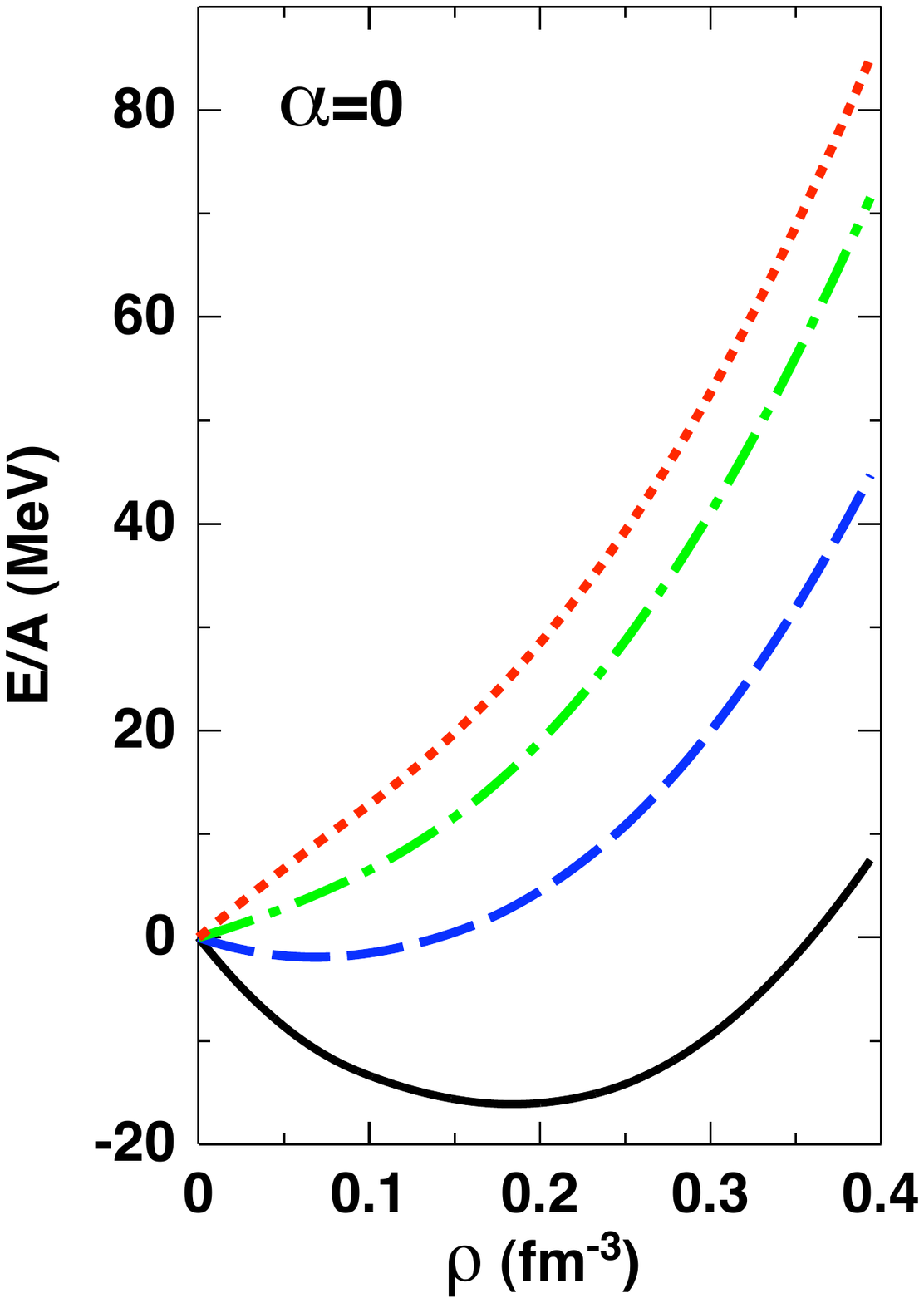}} 
\scalebox{0.25}{\includegraphics{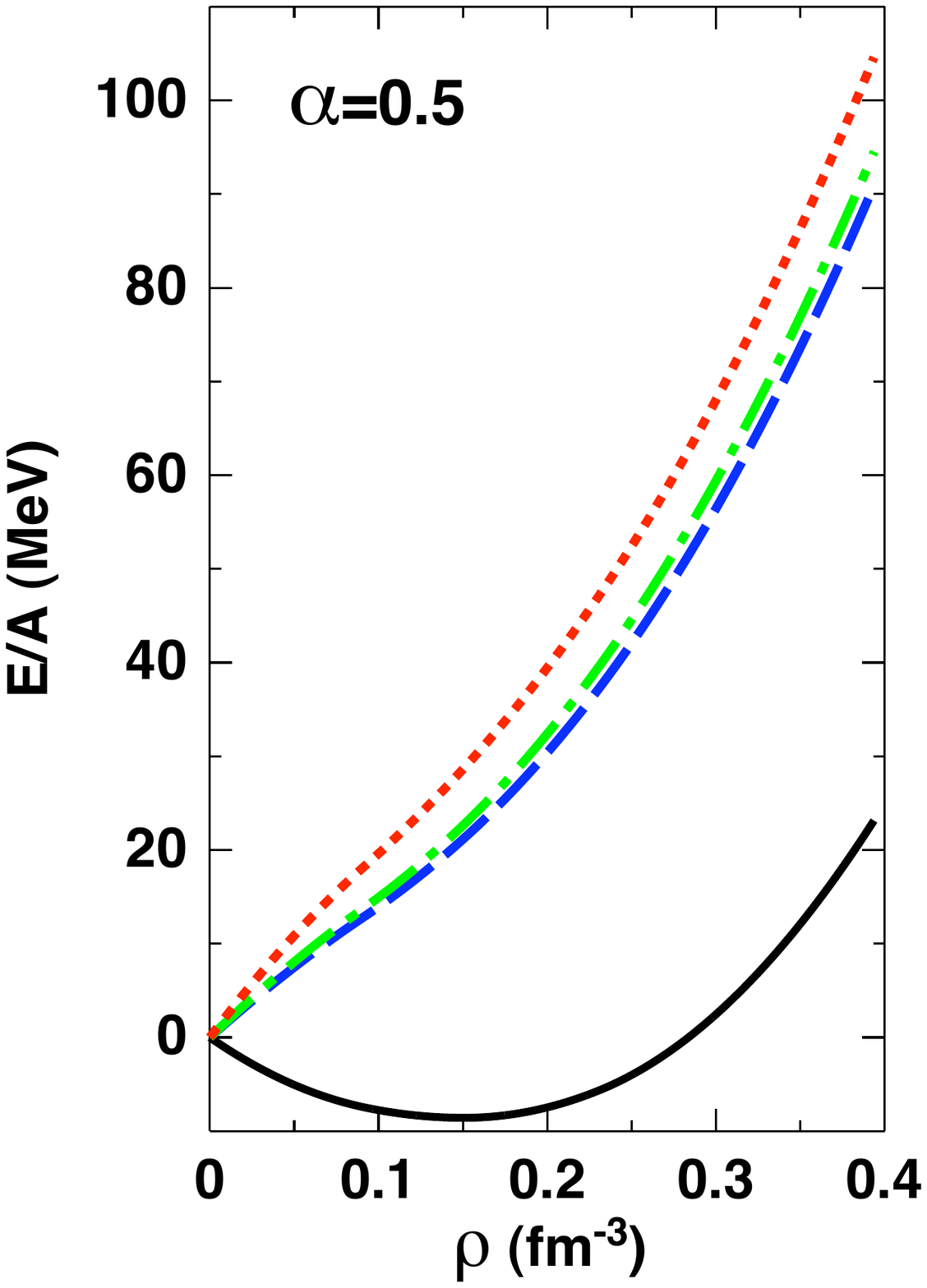}} 
\vspace*{-1.0cm}
\caption{                                        
The energy per particle as a function of density and variuos degrees of proton and neutron polarizations
in symmetric matter (left) and asymmetric matter (right). In both frames, the (blue) dashed line corresponds to 
totally polarized neutrons and unpolarized protons ($\beta_n$=1, $\beta_p$=0); the (green) dash-dotted line is the 
prediction for the FM state 
($\beta_n$=1, $\beta_p$=1); the (red) dotted line shows the energy of the AFM state 
($\beta_n$=1, $\beta_p$=-1).                                                         
The (black) solid line shows the predictions for unpolarized matter. 
} 
\label{siso}
\end{figure}

In the left panel of Fig.~\ref{siso}, we show, 
in comparison with unpolarized symmetric matter (solid line):            
the EoS 
for the case of fully polarized neutrons and completely unpolarized protons (dashed line); 
the EoS 
for the case of protons and neutrons totally polarized in the same direction, that is, matter in the 
ferromagnetic (FM) state ( dashed-dotted line); 
the EoS 
for the case of protons and neutrons totally polarized in opposite directions, namely matter in the antiferromagnetic 
(AFM) state ( dotted line). 
A similar comparison is shown in the right panel of Fig.~(\ref{siso}), but for isospin asymmetric matter. 
(Notice that all predictions  
are invariant under a global spin flip.)                                      

We conclude that, for both symmetric and asymmetric matter,                 
the energies of the FM and AFM states are higher than those of the corresponding unpolarized cases, with the 
AFM state being the most energetic. 
Thus, a phase transition to a spin-ordered state is not anticipated in our model.               
This conclusion seems to be shared by predictions of microscopic models, such as those based on conventional Brueckner-Hartree-Fock theory 
\citep{pol18}. On the other hand, calculations based on various parametrizations of Skyrme forces result in 
different conclusions. For instance, 
 with the {\it SLy4} and {\it SLy5} forces and the Fermi liquid 
formalism                                                                                     
  a phase transition  to the AFM state is predicted in asymmetric matter 
at a critical density equal to about 2-3 times normal density \citep{IY}.

In closing this brief section, it is interesting to remark that models based on 
realistic nucleon-nucleon potentials, whether relativistic or non-relativistic, are at least in qualitative 
agreement with one another in predicting 
more energy for totally polarized states (FM or AFM) 
 up to densities well above normal density.

\section{Summary and Conclusions} 
In this chapter, we have been concerned with the nuclear equation of state of 
isospin asymmetric nuclear matter, 
the main input for calculations of the properties of compact stars as well as
a variety of other systems, such as the neutron skin of neutron-rich nuclei.

After describing our microscopic approach to the development of the equation of state for
nuclear matter and neutron-rich matter, 
we presented a brief review 
of the structure equations leading to the prediction of neutron star properties. 
Microscopic predictions from different models employing three-body forces along
with the non-relativistic Brueckner-Hartree-Fock method have also been
shown for comparison. Large model dependence is seen among predictions, especially 
those involving the highest densities.

Rich and diverse effort is presently going on 
to improve the available constraints on the EoS or find new ones.                   
These constraints are usually extracted through the analysis of selected heavy-ion collision
observables.                    
At the same time, 
partnership between nuclear physics and astrophysics is becoming increasingly important towards 
advancing our understanding of exotic matter. 
The recently approved Facility for Rare Isotope Beams (FRIB), thanks to new powerful technical capabilities, 
will forge tighter links between the two disciplines, as it will allow access to rare isotopes 
which play a critical role in astrophysical processes but have not yet been observed in terrestrial 
laboratories.

\section{Acknowledgements} 
Support from the U.S. Department of Energy is gratefully acknowledged.
I am deeply indebted to Prof.~F. Weber for the use of his TOV code and many helpful advises and  
suggestions. I like to thank Mr.~Boyu Chen for help with the preparation of this manuscript. 
Some of the neutron star 
properties have been calculated using 
public software downloaded from the website {\it http://www.gravity.phys.uwm.edu/rns}.

\end{document}